\documentclass{vkacmconf}
\usepackage{epsfig}
\usepackage{latexsym}

\title{Roles Are Really Great!}

\PrepareText{Technical Report No. 822 \hfill
MIT Laboratory for Computer Science \hfill
November 2001}

\author{\Author{Viktor Kuncak, Patrick Lam, and Martin Rinard}\\
\Address{Laboratory for Computer Science\\
	Massachusetts Institute of Technology\\ 
	Cambridge, MA 02139} \\ 
	\Email{{\{vkuncak, plam, rinard\}@lcs.mit.edu}}}

\begin{document}

\maketitle

\renewcommand{\thefootnote}{\fnsymbol{footnote}}
\footnotetext[1]{
This research was supported in part by DARPA Contract
F33615-00-C-1692, NSF Grant CCR00-86154,
NSF Grant CCR00-63513, and an NSERC graduate scholarship.  \\
\noindent
}

\renewcommand{\thefootnote}{\arabic{footnote}}

\newtheorem{definition}{Definition}
\newtheorem{exa}[definition]{Example}
\newenvironment{example}{\begin{exa}\rm}
                        {\end{exa}}
\newtheorem{property}[definition]{Property}
\newtheorem{lemma}[definition]{Lemma}

\newcommand{\eop}{\vrule width4pt height4pt depth0pt}

\newcommand{\field}{\m{field}} 
\newcommand{\slot}{\m{slot}}   
\newcommand{\slotno}{\m{slotno}}
\newcommand{\nodes}{\m{nodes}} 
\newcommand{\nullRole}{\m{null}_R}
\newcommand{\identities}{\m{identities}}
\newcommand{\acyclic}{\m{acyclic}}
\newcommand{\rhoc}{\rho_c}
\newcommand{\locallyConsistent}{\m{locallyConsistent}}
\newcommand{\reduce}{\m{reduce}}
\newcommand{\roleExp}{R_{\m{exp}}}
\newcommand{\parammark}{\m{param}}

\newcommand{\Proc}{\m{Proc}}
\newcommand{\proclocal}{\m{local}}   
\newcommand{\procparam}{\m{param}} 
\newcommand{\procvar}{\m{var}}   
\newcommand{\prerole}{\m{preR}}  
\newcommand{\postrole}{\m{postR}} 
\newcommand{\paramnode}{\m{parnd}} 

\newcommand{\aliasNodes}{\m{aliases}} 
\newcommand{\accessible}{\m{accessible}} 

\newcommand{\Rcyc}{R_{\mbox{\sf\tiny CYC}}}
\newcommand{\Rinter}{R_{\mbox{\sf\tiny INTER}}}
\newcommand{\Rfinal}{R_{\mbox{\sf\tiny FINAL}}}
\newcommand{\Rinit}{R_{\mbox{\sf\tiny INIT}}}
\newcommand{\Rtree}{R_{\mbox{\sf\tiny TREE}}}

\newcommand{\lroot}{\m{proc}_i}

\newcommand{\RA}{{\cal RA}} 

\newcommand{\ra}{\mathop{\longrightarrow}}
\newcommand{\dra}{\mathop{\Longrightarrow}}

\newcommand{\rac}{\ra} 

\newcommand{\nullConst}{\vv{null}}
\newcommand{\Nat}{{\cal N}} 

\newcommand{\consistent}{\m{con}}
\newcommand{\consistentWith}{\m{conW}}

\newcommand{\CFG}{\m{CFG}}
\newcommand{\CFGE}{E_{\mbox{\sf \tiny CFG}}}
\newcommand{\CFGN}{N_{\mbox{\sf \tiny CFG}}}
\
\newcommand{\Gc}{G_c}
\newcommand{\Hc}{H_c}
\newcommand{\nullObj}{\m{null}_c}

\newcommand{\satisfied}{\m{satisfied}_c} 

\newcommand{\RRD}{\m{RRD}} 

\newcommand{\comment}[1]{\eqno \mbox{(#1)}}
\newcommand{\withp}[1]{\uplus \{#1\}}
\newcommand{\CFGEP}{\CFGE(\nte{proc})}
\newcommand{\atproc}{@\nte{proc}_i;s}

\newcommand{\nodesc}{\m{nodes}}
\newcommand{\onstagec}{\m{onstage}}
\newcommand{\offstagec}{\m{offstage}}

\newcommand{\onstage}{\m{onstage}}   
\newcommand{\offstage}{\m{offstage}} 

\newcommand{\nullAbstNode}{\m{null}} 
\newcommand{\abstproc}{\m{proc}} 
\newcommand{\concproc}{\m{proc}_c} 
\newcommand{\abstsatisfied}{\m{satisfied}}

\newcommand{\nil}{\m{null}}

\newcommand{\GS}{{\cal G}}
\newcommand{\Gb}{\bot_G}
\newcommand{\abstrel}{\mathop{\alpha}}
\newcommand{\conabstrel}{\mathop{\alpha_0}} 
\newcommand{\homo}{h}

\newcommand{\Formals}{\m{Formals}}
\newcommand{\Roots}{\m{Roots}}
\newcommand{\Auxs}{\m{Auxs}}

\newcommand{\RPE}{\m{RPE}}              
\newcommand{\OLDH}{\m{H}_{\m{init}}}  

\newcommand{\localCheck}{\m{localCheck}} 
\newcommand{\acycCheck}{\m{acycCheck}}
\newcommand{\nodeCheck}{\m{nodeCheck}}

\newcommand{\paramCheck}{\m{paramCheck}} 

\newcommand{\nte}[1]{\mbox{\sf #1}}     
\newcommand{\te}[1]{\mathop{\mbox{\tt"#1"}}} 
\newcommand{\opt}[1]{(#1)^{?}\>}  
\newcommand{\rep}[1]{(#1)^{*}\>}  

\newcommand{\vv}[1]{\mbox{\tt #1}}
\newcommand{\q}[1]{\mbox{\tt"#1"}\>}
\newcommand{\co}[1]{\mbox{#1}}

\newcommand{\tuple}[1]{\langle #1 \rangle}
\newcommand{\tu}[1]{\langle #1 \rangle}

\newcommand{\lefti}{[ \! [}
\newcommand{\righti}{] \! ]}
\newcommand{\tr}[1]{\lefti #1 \righti}
\newcommand{\str}[1]{\mathop{st}\lefti #1 \righti}
\newcommand{\ctr}[1]{\mathop{con}\lefti #1 \righti}

\newcommand{\m}[1]{\mbox{\sf #1}}

\newcommand{\logicalor}{~\vee~}
\newcommand{\logicaland}{~\wedge~}

\newcommand{\merge}{\mathrm{merge }}

\newtheorem{prop}{Proposition}
\newtheorem{cor}{Corollary}



\newcommand{\expandeq}{\mathop{\preceq}}
\newcommand{\expandeqwith}[1]{\mathop{\preceq}\limits^{#1}}
\newcommand{\contracteq}{\mathop{\succeq}}
\newcommand{\contracteqwith}[1]{\mathop{\succeq}\limits^{#1}}
\newcommand{\symexec}[1]{\mathop{\Longrightarrow}\limits^{\vv{\small #1}}}
\newcommand{\transrel}[1]{\mathop{\leadsto}\limits^{\vv{\small #1}}}
\newcommand{\assertonstage}[1]
 {\mathop{\longrightarrow}\limits^{{\small \m{onstage?(}}{#1}\m{)}}}
\newcommand{\instantiate}[2]{\mathop{\Uparrow}\limits_{#1}^{#2}}
\newcommand{\instantiatesome}{\Uparrow}
\newcommand{\summarize}[1]{\mathop{\Downarrow}\limits^{#1}}
\newcommand{\splt}[1]{\mathop{\|}\limits^{#1}}
\newcommand{\spltsome}{\|}

\newcommand{\expand}{\mathop{\prec}}
\newcommand{\expandn}{\mathop{\prec}\limits^{n}}
\newcommand{\expandnf}{\mathop{\prec}\limits^{n,f}}
\newcommand{\expandeqnf}{\mathop{\preceq}\limits^{n,f}}
\newcommand{\expandstar}{\mathop{\prec}\limits^{*}}
\newcommand{\contract}{\mathop{\succ}}
\newcommand{\contractn}{\mathop{\succ}\limits^{n}}
\newcommand{\contracteqn}{\mathop{\succeq}\limits^{n}}

\newcommand{\pow}[1]{2^{#1}}
\newcommand{\alimatch}[1]{{\mathop{\rhd}\limits_\rho}}
\newcommand{\lland}{\>\mbox{and}\>}
\newcommand{\llor}{\>\mbox{or}\>}
\newcommand{\lliff}{\>\mbox{iff}\>}
\newcommand{\meta}[1]{\footnote{[#1]}}

\newcommand{\vstack}[1]{\begin{array}{c} #1 \end{array}}
\newcommand{\ivstack}[1]{\begin{array}{rl} #1 \end{array}}
\newcommand{\rolesat}{\m{rolesat}}
\newcommand{\swing}{\m{swing}}
\newcommand{\normalize}{\m{normalize}}
\newcommand{\reach}{\m{reach}}
\newcommand{\cl}[1]{{#1}_{/\sim}}

\newcommand{\context}{\m{context}}
\newcommand{\effect}{\m{effect}}
\newcommand{\procread}{\m{read}}
\newcommand{\procmaywrite}{\m{mayWr}}
\newcommand{\procmustwrite}{\m{mustWr}}
\newcommand{\mustwriteno}{\m{mustWrNo}}

\newcommand{\Hi}{H_{\m{\tiny IC}}}
\newcommand{\Ki}{K_{\m{\tiny IC}}}
\newcommand{\rhoi}{\rho_{\m{\tiny IC}}}
\newcommand{\newnodes}{\m{newnodes}}
\newcommand{\allnodes}{\m{allnodes}}

\newcommand{\updateMustWrite}{\m{updateWr}}

\newcommand{\matchContext}{\m{matchContext}}
\newcommand{\match}{\m{match}}
\newcommand{\GM}{\Gamma}
\newcommand{\xchoose}{\m{ choose }}
\newcommand{\where}{\m{ where }}
\newcommand{\xfail}{\m{ fail }}
\newcommand{\xif}{\m{if }}
\newcommand{\xthen}{\m{ then }}
\newcommand{\xelse}{\m{else }}
\newcommand{\xelsif}{\m{ elsif }}
\newcommand{\return}{\m{ return }}
\newcommand{\candidates}{\m{candidates}}

\newcommand{\multiset}[1]{\{\!\{ #1 \}\!\}}

\newcommand{\RoleGraphs}{\m{RoleGraphs}}
\newcommand{\theseparams}{\m{paramnodes}}
\newcommand{\inaccessible}{\m{inaccessible}}
\newcommand{\notAffected}{\m{notAffected}}

\newcommand{\appeff}[1]{\mathop{\vdash}\limits^{#1}}

\newcommand{\doeffects}{\mathop{\longrightarrow}\limits^{\sf \tiny FX}}
\newcommand{\doroles}{\mathop{\longrightarrow}\limits^{\sf \tiny RR}}
\newcommand{\removeIfMustWrite}{\m{orem}}

\newcommand{\NEW}{\m{NEW}} 
\newcommand{\addNEW}{\m{addNEW}} 

\newcommand{\unknownRole}{\m{unknown}}
\newcommand{\errorheap}{\m{error}_c}

\newcommand{\transition}[2]{\begin{array}{ll} {#1} \ra \\ {#2}\end{array}}
\newcommand{\transitionc}[2]{\begin{array}{ll} {#1} \rac \\ {#2}\end{array}}

\newcommand{\rchange}{\m{roleChOk}} 

\newcommand{\singleton}{\m{singleton}} 

\newcommand{\GC}{\m{GC}} 

\newcommand{\cascadingOk}{\m{cascadingOk}} 
\newcommand{\neighbors}{\m{neighbors}} 

\newcommand{\multislots}{\m{multislots}}

\newcommand{\Scyc}{S_{\mathrm{cyc}}} 
\newcommand{\SR}{S_{\mathrm{R}}} 
\newcommand{\SNR}{S_{\mathrm{NR}}} 
\newcommand{\Hcyc}{H_{\mathrm{cyc}}} 
\newcommand{\Hoff}{H_{\mathrm{off}}} 
\newcommand{\HR}{H_{\mathrm{R}}} 
\newcommand{\HNR}{H_{\mathrm{NR}}} 
\newcommand{\hR}{h_{\mathrm{R}}} 
\newcommand{\hNR}{h_{\mathrm{NR}}} 

\newcommand{\BfNR}{B_{\mathrm{fNR}}}
\newcommand{\BtNR}{B_{\mathrm{tNR}}}
\newcommand{\BfR}{B_{\mathrm{fR}}}
\newcommand{\BtR}{B_{\mathrm{tR}}}

\newcommand{\AfNR}{A_{\mathrm{fNR}}}
\newcommand{\AtNR}{A_{\mathrm{tNR}}}
\newcommand{\AfR}{A_{\mathrm{fR}}}
\newcommand{\AtR}{A_{\mathrm{tR}}}

\newcommand{\Nf}{N_{\mathrm{f}}}
\newcommand{\Nt}{N_{\mathrm{t}}}

\newcommand{\acycCheckAll}{\m{acycCheckAll}}

\begin{abstract}
We present a new role system for specifying changing
referencing relationships of heap objects.  The role of an
object depends, in large part, on its aliasing relationships
with other objects, with the role of each object changing as
its aliasing relationships change.  Roles therefore capture
important object and data structure properties and provide
useful information about how the actions of the program
interact with these properties.  Our role system enables the
programmer to specify the legal aliasing relationships that
define the set of roles that objects may play, the roles of
procedure parameters and object fields, and the role changes
that procedures perform while manipulating objects.  We
present an interprocedural, compositional, and
context-sensitive role analysis algorithm that verifies that
a program respects the role constraints.
\end{abstract}

\tableofcontents

\section{Introduction}

Types capture important properties of the objects that
programs manipulate, increasing both the safety and
readability of the program.  Traditional type systems capture
properties (such as the format of data items stored in the
fields of the object) that are invariant over the lifetime
of the object.  But in many cases, properties that do change
are as important as properties that do not.  Recognizing the
benefit of capturing these changes, researchers have
developed systems in which the type of the object changes as
the values stored in its fields change or as the program
invokes operations on the
object~\cite{StromYemini86Typestate,
StromYellin93ExtendingTypestate,
DeLineFahndrich01EnforcingHighLevelProtocols,
XuETAL00SafetyCheckingPLDI,
XuETAL01TypestateCheckingMachineCode,
Chambers93PredicateClasses, GottlobETAL94Roles,
DrossopoulouETAL01Reclassification}.  These
systems integrate the concept of changing object states into
the type system.

The fundamental idea in this paper is that the state of each
object also depends on the data structures
in which it participates.  Our type system therefore
captures the referencing relationships that determine this
data structure participation.  As objects move between data
structures, their types change to reflect their changing
relationships with other objects.  Our system uses {\em
roles} to formalize the concept of a type that depends on
the referencing relationships.  Each role declaration provides
complete aliasing information for each object that plays
that role---in addition to specifying roles for the fields
of the object, the role declaration also identifies the
complete set of references in the heap that refer to the
object.  In this way roles generalize linear type systems
\cite{Wadler90LinearTypes, BarendsenSmetsers93UniquenessTyping, 
Kobayashi99QuasiLinearTypes} by allowing multiple aliases to
be statically tracked, and extend alias types
\cite{SmithETAL00AliasTypes,
WalkerMorrisett00AliasTypesRecursive} with the ability to
specify roles of objects that are the source of aliases.

This approach attacks a key difficulty associated with
state-based type systems: the need to ensure that any state
change performed using one alias is correctly reflected in
the declared types of the other aliases.  Because each
object's role identifies all of its heap aliases, the
analysis can verify the correctness of the role information
at all remaining or new heap aliases after an operation
changes the referencing relationships.

Roles capture important object and data structure
properties, improving both the safety and transparency of
the program.  For example, roles allow the programmer to
express data structure consistency properties (with the
properties verified by the role analysis), to improve the
precision of procedure interface specifications (by allowing
the programmer to specify the role of each parameter), to
express precise referencing and interaction behaviors
between objects (by specifying verified roles for object
fields and aliases), and to express constraints on the
coordinated movements of objects between data structures (by
using the aliasing information in role definitions to
identify legal data structure membership
combinations). Roles may also aid program optimization by
providing precise aliasing information.

This paper makes the following contributions:
\begin{itemize} \itemsep=0em
\item {\bf Role Concept:} The concept that the state of an
object depends on its referencing relationships; specifically,
that objects with different heap aliases should be regarded
as having different states. 
\item {\bf Role Definition Language:} It presents 
a language for defining roles. The programmer can use
this language to express data structure invariants and properties such as 
data structure participation.
\item {\bf Programming Model:}
It presents a set of role consistency
rules. These rules give a programming model for changing
the role of an object and the circumstances under which roles can
be temporarily violated.
\item {\bf Procedure Interface Specification Language:} 
It presents a language for specifying the initial context and
effects of each procedure. The effects summarize the actions
of the procedure in terms of the references it changes and
the regions of the heap that it affects.
\item {\bf Role Analysis Algorithm:} It presents an algorithm for
verifying that the program respects the constraints given by
a set of role definitions and procedure specifications.  The
algorithm uses a data-flow analysis to infer intermediate
referencing relationships between objects, allowing the
programmer to focus on role changes and procedure
interfaces.
\end{itemize}

\section{Example} \label{sec:example}

\begin{figure}
\begin{center}
\epsfig{file=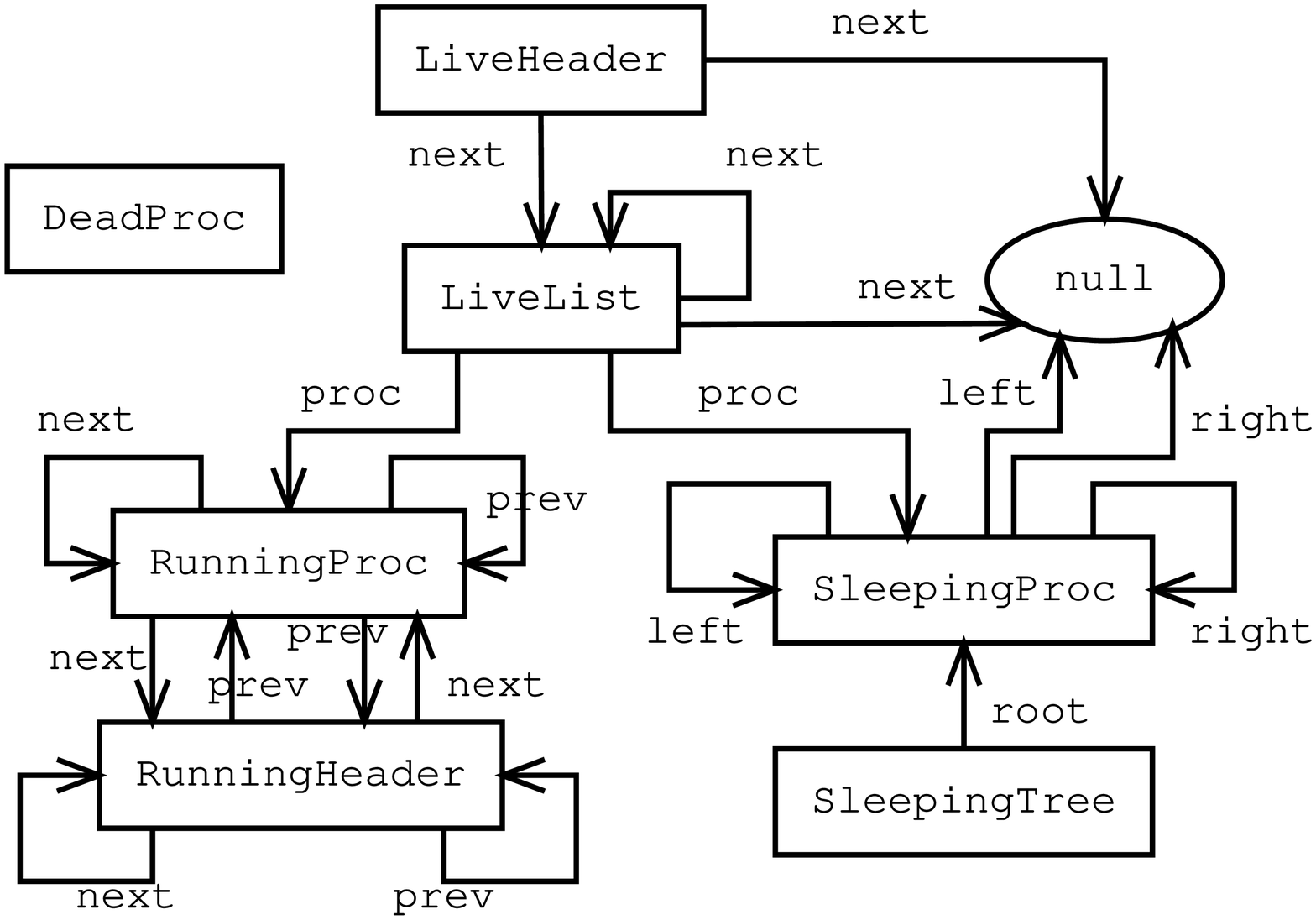,width=3in}
\end{center}
\caption{Role Reference Diagram for Scheduler}
\label{fig:schedulerSRD}
\end{figure}

Figure~\ref{fig:schedulerSRD} presents a {\em role reference
diagram} for a process scheduler.  Each box in the diagram
denotes a disjoint set of objects of a given role.  The
labelled arrows between boxes indicate possible references
between the objects in each set.  As the diagram indicates,
the scheduler maintains a list of live processes.
A live process can be either running or sleeping.  The
running processes form a doubly-linked list, while sleeping
processes form a binary tree.  Both kinds of processes have
\vv{proc} references from the live list nodes \vv{LiveList}.
Header objects \vv{RunningHeader} and \vv{SleepingTree}
simplify operations on the data structures that store the
process objects.

As Figure~\ref{fig:schedulerSRD} shows, data structure
participation determines the conceptual state of each
object.  In our example, processes that participate in the
sleeping process tree data structure are classified as
sleeping processes, while processes that participate in the
running process list data structure are classified as
running processes.  Moreover, movements between data
structures correspond to conceptual state changes---when a
process stops sleeping and starts running, it moves from the
sleeping process tree to the running process list.

\subsection{Role Definitions} 

Figure~\ref{fig:rolesScheduler} presents the role
definitions for the objects in our example.\footnote{In
general, each role definition would specify the static class
of objects that can play that role.  To simplify the
presentation, we assume that all objects are instances of a
single class with a set of fields $F$.}  Each role
definition specifies the constraints that an object must
satisfy to play the role.  Field constraints specify the
roles of the objects to which the fields refer, while slot
constraints identify the number and kind of aliases of the
object.

\begin{figure}[ht]
\begin{verbatim}
role LiveHeader {
  fields next : LiveList | null;
}
role LiveList {
  fields next : LiveList | null,
         proc : RunningProc | SleepingProc;
  slots  LiveList.next | LiveHeader.next;
  acyclic next;
}
role RunningHeader {
  fields next : RunningProc | RunningHeader,
         prev : RunningProc | RunningHeader;
  slots  RunningHeader.next | RunningProc.next,
         RunningHeader.prev | RunningProc.prev;
  identities next.prev, prev.next;
}
role RunningProc {
  fields next : RunningProc | RunningHeader,
         prev : RunningProc | RunningHeader;
  slots  RunningHeader.next | RunningProc.next,
         RunningHeader.prev | RunningProc.prev,
         LiveList.proc;
  identities next.prev, prev.next;
}
role SleepingTree {
  fields root : SleepingProc | null,
  acyclic left, right;
}
role SleepingProc {
  fields left  : SleepingProc | null,
         right : SleepingProc | null;
  slots SleepingProc.left | SleepingProc.right |
        SleepingTree.root;
        LiveList.proc;
  acyclic left, right;
}
role DeadProc {  }
\end{verbatim}
\caption{Role Definitions for a Scheduler}
\label{fig:rolesScheduler}
\end{figure}

Role definitions may also contain two additional kinds of
constraints: identity constraints, which specify paths that
lead back to the object, and acyclicity constraints, which
specify paths with no cycles. In our example, the identity
constraint \vv{next.prev} in the \vv{RunningProc} role
specifies the cyclic doubly-linked list constraint that
following the \vv{next}, then \vv{prev} fields always leads
back to the initial object.  The acyclic constraint
\vv{left, right} in the \vv{SleepingProc} role specifies
that there are no cycles in the heap involving only \vv{left} and
\vv{right} edges.  On the other hand, the list of running
processes must be cyclic because its nodes can never point
to \vv{null}.

The slot constraints specify the complete set of heap
aliases for the object.  In our example, this implies that
no process can be simultaneously running and sleeping.  

In general, roles can capture data structure consistency
properties such as disjointness and can prevent
representation exposure \cite{ClarkeETAL98OwnershipTypes}.
As a data structure description language, roles can
naturally specify trees with additional pointers.  Roles can
also approximate non-tree data structures like sparse
matrices.  Because most role constraints are local, it is
possible to inductively infer them from data structure
instances.

\subsection{Roles and Procedure Interfaces}

Procedures specify the initial and final roles of their
parameters.  The
\vv{suspend} procedure in Figure~\ref{fig:suspend}, for
example, takes two parameters: an object with role
\vv{RunningProc} \vv{p}, and the \vv{SleepingTree} \vv{s}.
The procedure changes the role of the object referenced by
\vv{p} to \vv{SleepingProc} whereas the object referenced by 
\vv{s} retains its original role.  To perform the role change, the procedure
removes \vv{p} from its \vv{RunningList} data structure and
inserts it into the \vv{SleepingTree} data structure
\vv{s}. If the procedure fails to perform the insertions or
deletions correctly, for instance by leaving an object in
both structures, the role analysis will report an error.

\begin{figure}[htbp]
\begin{verbatim}
procedure suspend(p : RunningProc ->> SleepingProc,
                  s : SleepingTree)
local pp, pn, r;
{
   pp = p.prev;   pn = p.next;
   r = s.root;
   p.prev = null; p.next = null;
   pp.next = pn;  pn.prev = pp;
   s.root = p;    p.left = r;
   setRole(p : SleepingProc);
}
\end{verbatim}
\caption{Suspend Procedure}
\label{fig:suspend}
\end{figure}

\section{Abstract Syntax and Semantics of Roles}

In this section, we precisely define what it means for a
given heap to satisfy a set of role definitions.  In
subsequent sections we will use this definition as a
starting point for a programming model and role analysis.

\subsection{Heap Representation}

We represent a concrete program heap as a finite directed
graph $\Hc$ with $\nodes(\Hc)$ representing objects of the
heap
and labelled edges representing heap references.  A graph
edge $\tu{o_1,f,o_2}\in \Hc$ denotes a reference with field
name $f$ from object $o_1$ to object $o_2$.  To simplify the
presentation, we fix a global set of fields $F$ and assume
that all objects have all fields in $F$.  We do not consider
subtyping or dynamic dispatch in this paper.

\subsection{Role Representation}

Let $R$ denote the set of roles used in role
definitions, $\nullRole$ be a special symbol always
denoting a null object $\nullObj$, and let $R_0 = R
\cup \{\nullRole\}$.  We represent each role as the conjunction
of the following four kinds of constraints:
\begin{itemize}
\item {\bf Fields:} 
For every field name $f \in F$ we
introduce a function $\field_f : R
\to 2^{R_0}$ denoting the set of roles that objects
of role $r \in R$ can reference through field $f$.  A field $f$ of
role $r$ can be null if and only if $\nullRole \in
\field_f(r)$.  The explicit use of $\nullRole$ and
the possibility to specify a set of alternative roles for every
field allows roles to express both may and must referencing
relationships.  
\item {\bf Slots:}
Every role $r$ has $\slotno(r)$ slots.  A
slot $\slot_k(r)$ of role $r \in R$ is a subset of $R \times F$.
Let $o$ be an object of role $r$ and $o'$ an object of role
$r'$.  A reference $\tu{o',f,o} \in \Hc$ can fill a slot $k$
of object $o$ if and only if $\tu{r',f} \in \slot_k(r)$.  An
object with role $r$ must have each of its slots filled by
exactly one reference.  
\item {\bf Identities:} Every role $r \in R$ has a set of
$\identities(r) \subseteq F \times F$.  Identities are pairs
of fields $\tu{f,g}$ such that following reference $f$ on
object $o$ and then returning on reference $g$ leads back to
$o$.
\item {\bf Acyclicities:} Every role $r \in R$ has a set $\acyclic(r)
\subseteq F$ of fields along which cycles are forbidden.
\end{itemize}

\subsection{Role Semantics} \label{sec:roleSemantics}

We define the semantics of roles as a conjunction of
invariants associated with role definitions.  A {\em
concrete role assignment} is a map $\rhoc : \nodesc(\Hc) \to
R_0$ such that $\rhoc(\nullObj) = \nullRole$.
\begin{definition} \label{def:heapConsistency}
Given a set of role definitions, we say that heap $\Hc$ is
{\em role consistent} iff there exists a role assignment
$\rhoc : \nodes(\Hc) \to R_0$ such that for every $o
\in \nodes(\Hc)$ the predicate $\locallyConsistent(o,\Hc,\rhoc)$
is satisfied.  We call any such role assignment $\rhoc$ a
{\em valid} role assignment.
\end{definition} 
The predicate $\locallyConsistent(o,\Hc,\rhoc)$ formalizes
the constraints associated with role definitions.
\begin{definition} \label{def:localConsistency}
$\locallyConsistent(o, \Hc,\rhoc)$ iff all of the following
conditions are met.  Let $r = \rhoc(o)$.
\begin{enumerate} \itemsep=0em

\item[1)] For every field $f \in F$ and $\tu{o,f,o'} \in \Hc$,
$\rhoc(o') \in \field_f(r)$.

\item[2)] Let $\{\tu{o_1,f_1},\ldots,\tu{o_k,f_k}\} = \{
\tu{o',f} \mid \tu{o',f,o} \in \Hc\}$ be the set of all aliases
of object $o$. Then $k = \slotno(r)$ and there exists some
permutation $p$ of the set $\{1,\ldots,k\}$ such
that $\tu{\rhoc(o_i),f_i} \in \slot_{p_i}(r)$ for all $i$.

\item[3)] If $\tu{o,f,o'} \in \Hc$, $\tu{o',g,o''} \in \Hc$,
      and \newline $\tu{f,g} \in \identities(r)$, then $o=o''$.

\item[4)] It is not the case that 
      graph $\Hc$ contains a cycle \newline
      $o_1,f_1,\ldots,o_s,f_s,o_1$ where
      $o_1 = o$ and \newline
      $f_1,\ldots,f_s \in \acyclic(r)$
\end{enumerate}
\end{definition}
Note that a role consistent heap may have multiple valid
role assignments $\rhoc$.  However, in each of these role
assignments, every object $o$ is assigned exactly one role
$\rhoc(o)$.  The existence of a role assignment $\rhoc$ with
the property $\rhoc(o_1) \neq \rhoc(o_2)$ thus implies
$o_1 \neq o_2$. This is just one of the ways in which roles make
aliasing more predictable.

\section{Role Properties}

Roles capture important properties of the objects and
provide useful information about how the actions of the
program affect those properties.
\begin{itemize}
\item {\bf Consistency Properties:} Roles can
ensure that the program respects application-level data structure consistency 
properties. The roles in our process scheduler, for example, ensure
that a process cannot be simultaneously sleeping and
running.  
\item {\bf Interface Changes:} In many cases, the interface of an 
object changes as its referencing relationships change. In
our process scheduler, for example, only running processes
can be suspended. Because procedures declare the roles of
their parameters, the role system can ensure that the
program uses objects correctly even as the object's
interface changes.
\item {\bf Multiple Uses:} Code factoring minimizes
code duplication by producing general-purpose
classes (such as the Java Vector and Hashtable classes) that can
be used in a variety of contexts. But this practice obscures the 
different purposes that different instances of these classes
serve in the computation. Because each instance's purpose is
usually reflected in its relationships with other objects, 
roles can often recapture these distinctions. 
\item {\bf Correlated Relationships:}
In many cases, groups of objects cooperate to implement a
piece of functionality. Standard type declarations provide
some information about these collaborations by identifying
the points-to relationships between related objects at the
granularity of classes. But roles can capture a much more
precise notion of cooperation, because they track
correlated state changes of related objects.  
\end{itemize}

Programmers can use roles for specifying the membership of
objects in data structures and the structural invariants of
data structures.  In both cases, the slot constraints are
essential.

When used to describe membership of an object in a data
structure, slots specify the source of the alias from a data
structure node that stores the object.  By assigning
different sets of roles to data structures used at different
program points, it is possible to distinguish nodes stored
in different data structure instances.  As an object moves
between data structures, the role of the object changes
appropriately to reflect the new source of the alias.

When describing nodes of data structures, slot constraints
specify the aliasing constraints of nodes; this is enough to
precisely describe a variety of data structures and
approximate many others.  Property~\ref{pro:treeness} below shows
how to identify trees in role definitions even if tree nodes
have additional aliases from other sets of nodes.  It is
also possible to define nodes which make up a compound data
structure linked via disjoint sets of fields, such as
threaded trees, sparse matrices and skip lists.

\begin{figure}[htb]
\begin{center}
\epsfig{file=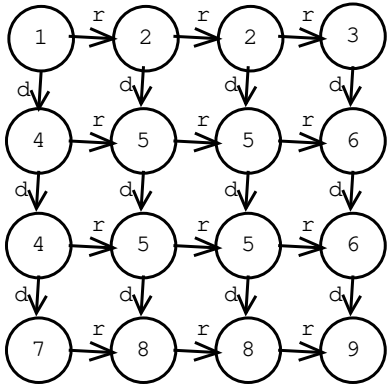,width=2in}
\end{center}
\caption{Roles of Nodes of a Sparse Matrix}
\label{fig:matrixSketch}
\end{figure}

\begin{example}
The following role definitions specify a sparse matrix
of width and height at least 3.  These definitions can be easily
constructed from a sketch of a sparse matrix, as in
Figure~\ref{fig:matrixSketch}.
\begin{verbatim}
role A1 {
  fields right : A2, down : A4;
  acyclic right, down;
}
role A2 {
  fields right : A2 | A3, down : A5;
  slots A1.right | A2.right;
  acyclic right, down;
}
role A3 {
  fields down : A6;
  slots A2.right;
  acyclic right, down;
}
role A4 {
  fields right : A5, down : A4 | A7;
  slots A1.down | A4.down;
  acyclic right, down;
}
role A5 {
  fields right : A5 | A6, down : A5 | A8;
  slots A4.right | A5.right, A2.down | A5.down;
  acyclic right, down;
}
role A6 {
  fields down : A6 | A9;
  slots A5.right, A3.down | A6.down;
  acyclic right, down;
}
role A7 {
  fields right : A8;
  slots A4.down;
  acyclic right, down;
}
role A8 {
  fields right : A8 | A9;
  slots A7.right | A8.right, A5.down;
  acyclic right, down;
}
role A9 {
  slots A8.right, A6.down;
  acyclic right, down;
}
\end{verbatim}
\end{example}

\begin{figure}
\begin{center}
\epsfig{file=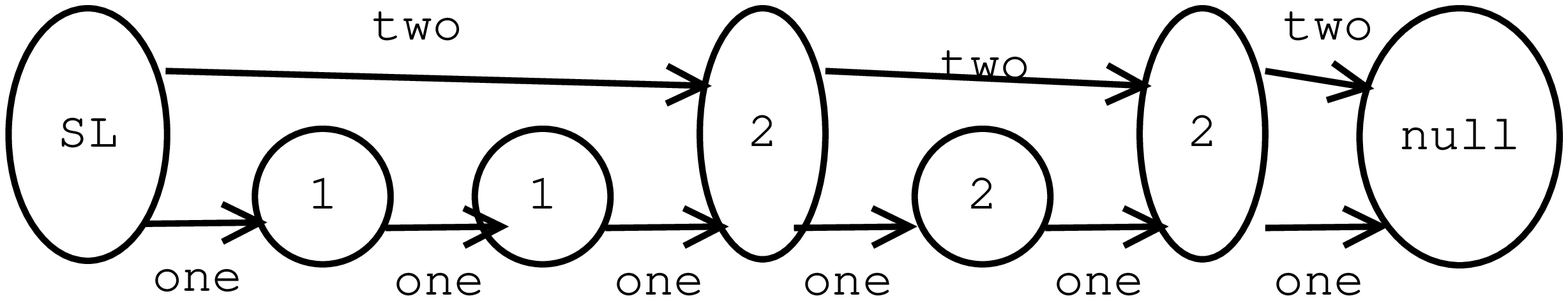,width=3in}
\end{center}
\caption{Sketch of a Two-Level Skip List}{}
\label{fig:skipListSketch}
\end{figure}

\begin{example}
We next give role definitions for a two-level skip list
\cite{Pugh90SkipList} sketched in Figure~\ref{fig:skipListSketch}.
\begin{verbatim}
role SkipList { 
  fields one : OneNode | TwoNode | null;
         two : TwoNode | null;
}
role OneNode {
  fields one : OneNode | TwoNode | null;
         two : null;
  slots OneNode.one | TwoNode.one | SkipList.one;
  acyclic one, two;
}
role TwoNode {
  fields one : OneNode | TwoNode | null;
         two : TwoNode | null;
  slots OneNode.one | TwoNode.one | SkipList.one,
        TwoNode.two | SkipList.two;
  acyclic one, two;
}
\end{verbatim}
\end{example}

\subsection{Formal Properties of Roles} \label{sec:properties}

In this section we identify some of the invariants
expressible using sets of mutually recursive role
definitions.  A further study of role properties can be
found in \cite{Kuncak01DesigningRoleAnalysis}.

The following properties show some of the ways role
specifications make object aliasing more predictable.  They
are an immediate consequence of the semantics of roles.
\begin{property}(Role Disjointness) \\
If there exists a valid role assignment $\rhoc$ for $\Hc$
such that $\rho(o_1) \neq \rho(o_2)$,
then $o_1 \neq o_2$.
\end{property}
The previous property gives a simple criterion for showing
that objects $o_1$ and $o_2$ are unaliased: find a valid
role assignment which assigns different roles to $o_1$ and
$o_2$.  This use of roles generalizes the use of static
types for pointer analysis
\cite{DiwanETAL98TypeBasedAliasAnalysis}.
Since roles create a finer partition of objects than a
typical static type system, their potential for proving
absence of aliasing is even larger.

\begin{property}(Disjointness Propagation) \\
If $\tu{o_1,f,o_2}, \tu{o_3,g,o_4} \in \Hc$, $o_1 \neq o_3$,
and there exists a valid role assignment $\rhoc$ for $\Hc$ such that
$\rhoc(o_2) = \rhoc(o_4) = r$ but
$\field_f(r) \cap \field_g(r) = \emptyset$,
then $o_2 \neq o_4$.
\end{property}
\begin{property}(Generalized Uniqueness) \label{pro:genLin} \\
If $\tu{o_1,f,o_2}, \tu{o_3,g,o_4} \in \Hc$, $o_1 \neq o_3$,
and there exists a role assignment $\rhoc$ such that
$\rhoc(o_2) = \rhoc(o_4) = r$, but
there are no indices $i \neq j$ such
that $\tu{\rhoc(o_1),f} \in \slot_i(r)$
and $\tu{\rhoc(o_2),g} \in \slot_j(r)$
then $o_2 \neq o_4$.
\end{property}
A special case of Property~\ref{pro:genLin} occurs when
$\slotno(r) = 1$; this constrains all references to objects of
role $r$ to be unique.

Role definitions induce a role reference diagram $\RRD$ which 
captures some, but not all, role constraints.
\begin{definition}(Role Reference Diagram) \label{def:RRD} \\
Given a set of definitions of roles $R$, a
role reference diagram $\RRD$ is is a directed graph with 
nodes $R_0$ and labelled edges defined by
\[\begin{array}{r@{}l}
   \RRD {} = \ & \{ \tu{r,f,r'} \mid
      r' \in \field_f(r) \mbox{ and }
      \exists i \>\> \tu{r,f} \in \slot_i(r') \} \\
          &\ {} \cup \{ \tu{r,f,\nullRole} \mid
      \nullRole \in \field_f(r) \}
\end{array}\]
\end{definition}
Each role reference diagram is a refinement of the
corresponding class diagram in a statically typed
language, because it partitions classes into multiple roles
according to their referencing relationships.  The sets
$\rhoc^{-1}(r)$ of objects with role $r$ change during
program execution, reflecting the changing referencing
relationships of objects.

Role definitions give more information than a role reference
diagram.  Slot constraints specify not only that objects of
role $r_1$ can reference objects of role $r_2$ along field
$f$, but also give cardinalities on the number of references
from other objects.  In addition, role definitions include
identity and acyclicity constraints, which are not present
in role reference diagrams.

\begin{property} \label{pro:RRD}
Let $\rhoc$ be any valid role assignment.  Define
\[
   G = \{ \tu{\rhoc(o_1),f,\rhoc(o_2)} \mid 
            \tu{o_1,f,o_2} \in \Hc \}
\]
Then $G$ is a subgraph of $\RRD$.
\end{property}
It follows from Property~\ref{pro:RRD} that roles give an
approximation of may-reachability among heap objects.
\begin{property}(May Reachability) \\
If there is a valid role assignment $\rhoc : \nodesc(\Hc)
\to R_0$
such that $\rhoc(o_1) \neq \rhoc(o_2)$ where $o_1, o_2 \in
\nodesc(\Hc)$ and there is no path from $\rhoc(o_1)$ to
$\rhoc(o_2)$ in the role reference diagram $\RRD$, then
there is no path from $o_1$ to $o_2$ in $\Hc$.
\end{property}
The next property shows the advantage of explicitly
specifying null references in role definitions.  While the
ability to specify acyclicity is provided by the
$\vv{acyclic}$ constraint, it is also possible to indirectly
specify must-cyclicity.
\begin{property}(Must Cyclicity) \\
Let $F_0 \subseteq F$ and $\Rcyc \subseteq R$ be a set of
nodes in the role reference diagram $\RRD$ such that for every
node $r \in \Rcyc$, if $\tu{r,f,r'} \in
\RRD$ then $r' \in \Rcyc$.  If $\rhoc$ is a valid role
assignment for $\Hc$, then every object $o_1 \in \Hc$ with
$\rhoc(o_1) \in \Rcyc$ is a member of a cycle in $\Hc$ with
edges from $F_0$.
\end{property}
The following property shows that roles can specify a form
of must-reachability among the sets of objects with the same
role.
\begin{property}(Downstream Path Termination) \\ \label{pro:FWDtermination}
Assume that for some set of fields $F_0 \subseteq F$ there
are sets of nodes $\Rinter \subseteq R$, $\Rfinal \subseteq
R_0$ of the role reference diagram $\RRD$ such that for every
node $r \in \Rinter$:
\begin{enumerate}
\item $F_0 \subseteq \acyclic(r)$
\item if $\tu{r,f,r'} \in \RRD$ for $f \in F_0$,
      then $r' \in \Rinter \cup  \Rfinal$
\end{enumerate}
Let $\rhoc$ be a valid role assignment for $\Hc$.  Then
every path in $\Hc$ starting from an object $o_1$ with role
$\rhoc(o_1) \in \Rinter$ and containing only edges labelled
with $F_0$ is a prefix of a path that terminates at some
object $o_2$ with $\rhoc(o_2) \in \Rfinal$.
\end{property}

\begin{property}(Upstream Path Termination) \\ \label{pro:BACKtermination}
Assume that for some set of fields $F_0 \subseteq F$ there
are sets of nodes $\Rinter \subseteq R$, $\Rinit \subseteq
R_0$ of the role reference diagram $\RRD$ such that for every
node $r \in \Rinter$:
\begin{enumerate}
\item $F_0 \subseteq \acyclic(r)$
\item if $\tu{r',f,r} \in \RRD$ for $f \in F_0$,
      then $r' \in \Rinter \cup  \Rinit$
\end{enumerate}
Let $\rhoc$ be a valid role assignment for $\Hc$.  Then
every path in $\Hc$ terminating at an object $o_2$ with
$\rhoc(o_2) \in \Rinter$ and containing only edges labelled
with $F_0$ is a suffix of a path which started at some
object $o_1$, where $\rhoc(o_1) \in \Rinit$.
\end{property}
The next two properties guarantee reachability properties 
by which there must exist at
least one path in the heap, rather than stating properties
of all paths as in Properties
\ref{pro:FWDtermination} and~\ref{pro:BACKtermination}.
\begin{property}(Downstream Must Reachability) \\
\label{pro:FWDreach}
Assume that for some set of fields $F_0 \subseteq F$ there
are sets of roles $\Rinter \subseteq R$, $\Rfinal \subset
R_0$ of the role reference diagram $\RRD$ such that for every
node $r \in \Rinter$:
\begin{enumerate}
\item $F_0 \subseteq \acyclic(r)$
\item there exists $f \in F_0$ such that 
      $\field_f(r) \subseteq \Rinter \cup \Rfinal$
\end{enumerate}
Let $\rhoc$ be a valid role assignment for $\Hc$.  Then for
every object $o_1$ with $\rhoc(o_1) \in \Rinter$ there is a
path in $\Hc$ with edges from $F_0$ from $o_1$ to some
object $o_2$ where $\rhoc(o_2) \in \Rfinal$.
\end{property}

\begin{property}(Upstream Must Reachability) \\
\label{pro:BACKreach}
Assume that for some set of fields $F_0 \subseteq F$ there
are sets of nodes $\Rinter \subseteq R$, $\Rinit \subseteq
R$ of the role reference diagram $\RRD$ such that for every
node $r \in \Rinter$:
\begin{enumerate}
\item $F_0 \subseteq \acyclic(r)$
\item there exists $k$ such that 
      $\slot_k(r) \subseteq (\Rinter \cup \Rinit)\times F$
\end{enumerate}
Let $\rhoc$ be a valid role assignment for $\Hc$.  Then for
every object $o_2$ with $\rhoc(o_2) \in \Rinter$ there is a
path in $\Hc$ from some object $o_1$ with $\rhoc(o_1) \in
\Rinit$ to the object $o_2$.
\end{property}


Trees are a class of data structures especially suited for
static analysis.  Roles can express graphs that are not
trees, but it is useful to identify trees as certain sets of
mutually recursive role definitions.
\begin{property}(Treeness) \\ \label{pro:treeness}
Let $\Rtree \subseteq R$ be a set of roles and $F_0
\subseteq F$ set of fields such that for every $r \in \Rtree$
\begin{enumerate}
\item $F_0 \subseteq \acyclic(r)$
\item $|\{i \mid \slot_i(r) \cap (\Rtree \times F_0) \neq \emptyset \}| \leq 1$
\end{enumerate}
Let $\rhoc$ be a valid role assignment for $\Hc$ and
$S \subseteq \{ \tu{n_1,f,n_2} \mid \tu{n_1,f,n_2} \in \Hc,
\rho(n_1),\rho(n_2) \in \Rtree, f \in F_0 \}$.
Then $S$ is a set of trees.
\end{property}

%
%







\section{A Programming Model} \label{sec:progMod}

In this section we define what it means for an execution of
a program to respect the role constraints. This definition
is complicated by the need to allow the program to
temporarily violate the role constraints during data
structure manipulations. Our approach is to let the program
violate the constraints for objects referenced by local
variables or parameters, but require all other objects to
satisfy the constraints.

We first present a simple imperative language with dynamic
object allocation and give its operational semantics.  We
then specify additional statement preconditions that enforce
the role consistency requirements.

\subsection{A Simple Imperative Language}

Our core language contains, as basic statements, Load
(\vv{x=y.f}), Store (\vv{x.f=y}), Copy (\vv{x=y}), and New
(\vv{x=new}).  All variables are references to objects in
the global heap and all assignments are reference
assignments.  We use an elementary \vv{test} statement
combined with nondeterministic choice and iteration to
express \vv{if} and \vv{while} statement, using the usual
translation \cite{HarelETAL00DynamicLogic,
BallETAL01AutomaticPredicateAbstraction}.  We represent the
control flow of programs using control-flow graphs.

A program is a collection of procedures $\m{proc} \in
\Proc$.  Procedures change the global heap but do not return values.
Every procedure $\m{proc}$ has a list of parameters
$\procparam(\m{proc}) = \{ \procparam_i(\m{proc})
\}_i$ and a list of local variables $\proclocal(\m{proc})$.
We use $\procvar(\m{proc})$ to denote $\procparam(\m{proc}) \cup
\proclocal(\m{proc})$.
A procedure definition specifies the initial role
$\prerole_k(\m{proc})$ and the final role
$\postrole_k(\m{proc})$ for every parameter
$\procparam_k(\m{proc})$.  We use $\m{proc}_j$ for indices
$j
\in \Nat$ to denote activation records of procedure
$\m{proc}$.  We further assume that there are no
modifications of parameter variables so every parameter
references the same object throughout the lifetime of
procedure activation.

\begin{example} \label{exa:processOp}
The following \vv{kill} procedure removes a process from both the
doubly linked list of running processes and the list of all
active processes.  This is indicated by the transition
from \vv{RunningProc} to \vv{DeadProc}.

\begin{verbatim}
procedure kill(p : RunningProc ->> DeadProc, 
               l : LiveHeader)
local prev, current, cp, nxt, lp, ln;
{
  // find 'p' in 'l'
  prev = l; current = l.next;
  cp = current.proc;
  while (cp != p) {
    prev = current;
    current = current.next;
    cp = current.proc;
  }
  // remove 'current' and 'p' from active list
  nxt = current.next; 
  prev.next = nxt; current.
  current.proc = null;
  setRole(current : IsolatedCell);
  // remove 'p' from running list
  lp = p.prev;   ln = p.next;
  p.prev = null; p.next = null;
  lp.next = ln;   ln.prev = lp;
  setRole(p : DeadProc);
}
\end{verbatim}
\end{example}

\subsection{Operational Semantics} \label{sec:operatSem}


\begin{figure*}


\begin{squeeze}
\[\begin{array}{|c|l|c|c|c} \hline
\mbox{Statement} &
\mbox{Transition} &
\mbox{Constraints} &
\mbox{Role Consistency} \\ \hline \hline
\vstack{
        p : \vv{x=y.f}
       }                                                        &
\transition{\tu{p\atproc,\Hc \withp{\tu{\lroot,\vv{x},o_x}}}}
           {\tu{p'\atproc,\Hc'}}                                &
\vstack{\vv{x}, \vv{y} \in \proclocal(\m{proc}), \\
       \tu{\lroot,\vv{y},o_y}, \tu{o_y,\vv{f},o_f} \in \Hc, \\
       \tu{p,p'} \in \CFGEP, \\
       \Hc' = \Hc \withp{\lroot,\vv{x},o_f}
       }                                                        &
\vstack{
        \accessible(o_f, \lroot, \Hc), \\
        \consistent(\Hc',\offstagec(\Hc')) 
       }                                                        \\ \hline
\vstack{
        p: \vv{x.f=y}
       }                                                        &
\transition{\tu{p\atproc,\Hc \withp{\tu{o_x,f,o_f}}}}
           {\tu{p'\atproc,\Hc'}}                                &
\vstack{\vv{x}, \vv{y} \in \proclocal(\m{proc}), \\
        \tu{\lroot,\vv{x},o_x}, \tu{\lroot,\vv{y},o_y} \in \Hc, \\
        \tu{p,p'} \in \CFGEP, \\
        \Hc' = \Hc \withp{\tu{o_x,f,o_y}}
       }                                                        &
\vstack{o_f \in \onstagec(\Hc,\m{proc}_i) \\
        \consistent(\Hc',\offstagec(\Hc'))
       }                                                        \\ \hline
p: \vv{x=y}                                                   &
\transition{\tu{p\atproc,\Hc \withp{\tu{\lroot,\vv{x},o_x}}}}
           {\tu{p'\atproc,\Hc'}}       &
\vstack{\vv{x} \in \proclocal(\m{proc}), \\
        \vv{y} \in \procvar(\m{proc}), \\
        \tu{\lroot,\vv{y},o_y} \in \Hc, \\
        \tu{p,p'} \in \CFGEP, \\
        \Hc' = \Hc \withp{\tu{\lroot,\vv{x},o_y}}
       }                                                        &
\vstack{
        \consistent(\Hc',\offstagec(\Hc'))
       }                                                        \\ \hline
p: \vv{x=new}                                                 &
\transition{\tu{p\atproc,\Hc \withp{\tu{\lroot,\vv{x},o_x}}}}
           {\tu{p'\atproc,\Hc'}}                                &
\vstack{\vv{x} \in \proclocal(\m{proc}), \\
        o_n \mbox{ fresh}, \\
        \tu{p,p'} \in \CFGEP, \\
        \Hc' = \Hc \withp{\tu{\lroot,\vv{x},o_n}} \uplus \m{nulls}, \\
        \m{nulls} = \{o_n\}\times F \times \{\m{null}\}
       }                                                        &
\vstack{
        \consistent(\Hc',\offstagec(\Hc'))
       }                                                        \\ \hline
p: \vv{test(c)}                                                 &
\transition{\tu{p\atproc, \Hc}}
           {\tu{p'\atproc, \Hc}}                                &
\vstack{\satisfied(\vv{c}, \lroot,\Hc), \\
        \tu{p,p'} \in \CFGEP}                                   &
\vstack{
        \consistent(\Hc,\offstagec(\Hc))
       }                                                        \\ \hline
\end{array}\]
\[  \satisfied(\vv{x==y}, \lroot, \Hc) \mbox{ iff } 
       \{ o \mid \tu{\lroot,\vv{x},o} \in \Hc \} =
       \{ o \mid \tu{\lroot,\vv{y},o} \in \Hc \}
\]
\[  \satisfied(\vv{!(x==y)}, \lroot, \Hc) \mbox{ iff not }
       \satisfied(\vv{x==y}, \lroot, \Hc)
\]

\[\begin{array}{rcl}
   \accessible(o, \lroot, \Hc) & := & 
       (\exists p \in \procparam(\m{proc}) : \tu{\lroot,p,o} \in \Hc) \\
                             &  & \mbox{ or } 
       \mbox{ not } (\exists \m{proc}'_j \>\>
              \exists v \in \procvar(\m{proc}') :
              \tu{\m{proc}'_j,v,o} \in \Hc)
\end{array}\]
\end{squeeze}

\caption{Semantics of Basic Statements}
\label{fig:operationalSemantics} 
\end{figure*}

In this section we give the operational semantics for our
language.  We focus on the first three columns in Figures
\ref{fig:operationalSemantics} and
\ref{fig:interProcOpSem}; the safety conditions in the
fourth column are detailed in Section~\ref{sec:safetyCond}.

Figure~\ref{fig:operationalSemantics} gives the small-step
operational semantics for the basic statements.  We use $A
\uplus B$ to denote the union $A \cup B$ where the sets
$A$ and $B$ are disjoint.  The program state consists of the
stack $s$ and the concrete heap $\Hc$.  The stack $s$ is a
sequence of pairs $p@\m{proc}_i \in
\times (\Proc \times \Nat)$, where $p \in \CFGN(\m{proc})$ is a program
point, and $\m{proc}_i \in \Proc \times \Nat$ is an
activation record of procedure $\m{proc}$.  Program points
$p \in \CFGN(\m{proc})$ are nodes of the control-flow
graphs. There is one control-flow graph for every procedure
$\m{proc}$.  An edge of the control-flow graph $\tu{p,p'}\in
\CFGE(\m{proc})$ indicates that control may transfer
from point $p$ to point $p'$.  We write $p : \m{stat}$ to
state that program point $p$ contains a statement
$\m{stat}$.  The control flow graph of each procedure contains special
program points
\vv{entry} and \vv{exit} indicating procedure entry and
exit, with no statements associated with them.  We assume
that all conditions are of the form \vv{x==y} or
\vv{!(x==y)} where \vv{x} and
\vv{y} are either variables or a special constant $\nullConst$
which always points to the $\nullObj$ object.

The concrete heap is either an error heap $\errorheap$ or a
non-error heap.  A non-error heap $\Hc \subseteq N \times F
\times N \cup ((\Proc
\times \Nat) \times V \times N)$ is a directed graph with
labelled edges, where nodes represent objects and procedure
activation records, whereas edges represent heap references
and local variables.  
An edge $\tu{o_1,f,o_2}
\in N \times F \times N$ denotes a reference from object
$o_1$ to object $o_2$ via field $f \in F$.  An edge
$\tu{\m{proc}_i,\vv{x},o} \in \Hc$ means that local variable
$\vv{x}$ in activation record $\m{proc}_i$ points to object
$o$.

A load statement $\vv{x=y.f}$ makes the variable $\vv{x}$
point to node $o_f$, which is referenced by the $\vv{f}$
field of object $o_y$, which is in turn referenced by
variable $\vv{y}$.  A store statement $\vv{x.f=y}$ replaces
the reference along field $\vv{f}$ in object $o_x$ by a
reference to object $o_y$ that is referenced by $\vv{y}$.
The copy statement $\vv{x=y}$ copies a reference to object
$o_y$ into variable $\vv{x}$.  The statement $\vv{x=new}$
creates a new object $o_n$ with all fields initially
referencing $\nullObj$, and makes $\vv{x}$ point to $o_n$.
The statement $\vv{test(c)}$ allows execution to proceed
only if condition $\vv{c}$ is satisfied.

\begin{figure*}

\begin{squeeze}
\[\begin{array}{|c|l|c|c|c} \hline
\mbox{Statement} &
\mbox{Transition} &
\mbox{Constraints} &
\mbox{Role Consistency} \\ \hline \hline

\vv{entry} : \_                                                         &
\transition{\tu{p@\m{proc}_i;s, \Hc}}
           {\tu{p'@\m{proc}_i;s, \Hc \uplus \m{nulls}}}                 &
\ivstack{
         \m{nulls} & = \{
          \tu{\m{proc}_i,v,\nullObj} \mid \\
           & v \in \proclocal(\m{proc}), \\
         \tu{p,p'} & \in \CFGEP
       }                                                                &
\vstack{
        \consistent(\Hc, \offstagec(\Hc))
       }                                                                \\ \hline
p : \m{proc}'(x_k)_k                                                    &
\transition{\tu{p@\m{proc}_i;s,\Hc}}
           {\tu{{\tt entry}@\m{proc}'_j;p'@\m{proc}_i;s,
           \Hc'}}                                                       &
\vstack{
        j \mbox{ fresh in } p@\m{proc}i;s, \\
        \tu{p,p'} \in \CFGEP, \\
        o_k : \tu{\m{proc}_i,x_k,o_k} \in \Hc, \\        
        \Hc' = \Hc \uplus \{ \tu{\m{proc}'_j,p_k,o_k} \}_k,\\
        \forall k \>\> p_k = \procparam_k(\m{proc}')
       }                                                                &
\vstack{
        \consistentWith(\m{ra},\Hc, S), \\
        \m{ra} = \{ \tu{o_k,\prerole_k(\m{proc}')} \}_k, \\
        S = \offstagec(\Hc) \cup \{ o_k \}_k
       }                                                                \\ \hline
\vv{exit} : \_                                                          &
\transition{\tu{p@\m{proc}_i;s, \Hc}}
           {\tu{s, \Hc \setminus \m{AF}}}                               &
\vstack{
        \m{AF} = \{ \tu{\m{proc}_i,v,n} \mid \\
           \tu{\m{proc}_i,v,n} \in \Hc \}
       }                                                                &
\vstack{
        \consistentWith(\m{ra},\Hc, S), \\
        \m{ra} = \{\tu{\paramnode_k(\m{proc}_i),\postrole_k(\m{proc})} \}_k, \\
        \ivstack{ S = &\offstagec(\Hc) \cup {} \\
                    &   \{ o \mid \tu{\m{proc}_i,v,o} \in \Hc \}}
       }                                                                \\ \hline
\end{array}\]

\[
      \paramnode_k(\m{proc}_i) = o
  \mbox{ where }
      \tu{\m{proc}_i,\procparam_k(\m{proc}),o} \in \Hc
\]
\end{squeeze}

\caption{Semantics of Procedure Call}
\label{fig:interProcOpSem}
\end{figure*}

Figure~\ref{fig:interProcOpSem} describes the semantics of
procedure calls.  Procedure call pushes new activation
record onto stack, inserts it into the heap, and initializes
the parameters.  Procedure entry initializes local
variables.  Procedure exit removes the activation record
from the heap and the stack.

\subsection{Onstage and Offstage Objects} \label{sec:onstageOffstage}

At every program point the set of all objects of heap $\Hc$ can be partitioned into:
\begin{enumerate} \itemsep=0em
\item {\bf onstage objects} ($\onstagec(\Hc)$) referenced by a local variable or parameter
of some activation frame;
\[\begin{array}{r@{}c@{}l}
  \onstagec(\Hc,\m{proc}_i) &:= & \{ o \mid  
           \exists x \in \procvar(\m{proc}) \\
       & &\quad      \tu{\m{proc}_i,x,o} \in \Hc \} \\ 
  \onstagec(\Hc) &:= & \bigcup\limits_{\m{proc}_i} \onstagec(\Hc,\m{proc}_i)
\end{array}\]
\item {\bf offstage objects} ($\offstagec(\Hc)$) unreferenced by local or
           parameter variables.
\[
   \offstagec(\Hc) := \nodesc(\Hc) \setminus \onstagec(\Hc)
\]
\end{enumerate}
Onstage objects need not have correct roles.  Offstage
objects must have correct roles assuming some role
assignment for onstage objects; the exception is that
acyclicity constraints for offstage objects can be violated
due to cycles that pass through the onstage objects.
\begin{definition} \label{def:heapUptoConsistency}
Given a set of role definitions and a set of objects $S_c
\subseteq \nodesc(S_c)$, we
say that heap $\Hc$ is {\em role consistent for $S_c$}, and
we write $\consistent(\Hc,S_c)$, iff there exists a role
assignment $\rhoc : \nodes(\Hc) \to R_0$ such that the
predicate $\locallyConsistent(o,\Hc,\rhoc,S_c)$ is satisfied
for every object $o \in S_c$.
\end{definition}
We define $\locallyConsistent(o,\Hc,\rhoc,S_c)$ to
generalize the $\locallyConsistent(o,\Hc,\rhoc)$ predicate, weakening
the acyclicity condition.
\begin{definition} \label{def:localUptoConsistency}
$\locallyConsistent(o, \Hc,\rhoc, S_c)$ holds iff conditions
1), 2), and 3)
of Definition~\ref{def:localConsistency}
are satisfied and the following condition holds:
\begin{enumerate} \itemsep=0em
\item[4')] It is not the case that 
      graph $\Hc$ contains a cycle \newline
      $o_1,f_1,\ldots,o_s,f_s,o_1$ such that \newline
      $o_1 = o$, 
      $f_1,\ldots,f_s \in \acyclic(r)$, and \newline additionally
      $o_1,\ldots,o_s \in S_c$.
\end{enumerate}
\end{definition}
Here $S_c$ is the set of onstage objects that are not allowed
to create a cycle; objects in $\nodesc(\Hc) \setminus S_c$
are exempt from the acyclicity condition.
The $\locallyConsistent(o, \Hc,\rhoc, S_c)$ and
$\consistent(\Hc,S_c)$ predicates are monotonic in $S_c$, so
a larger $S_c$ implies a stronger invariant.  
For $S_c =
\nodesc(\Hc)$, consistency for $S_c$ is equivalent with heap
consistency from Definition~\ref{def:heapConsistency}.  Note
that the role assignment $\rhoc$ specifies roles even for
objects $o
\in \nodesc(\Hc) \setminus S_c$.  This is because the role of $o$
may influence the role consistency of objects in $S_c$ which
are adjacent to $o$.

At procedure calls, the role declarations for parameters
restrict the set of potential role assignments. We 
therefore generalize $\consistent(\Hc,S_c)$ to $\consistentWith(\m{ra},\Hc,S_c)$,
which restricts the set of role assignments $\rhoc$
considered for heap consistency.
\begin{definition}
Given a set of role definitions, a heap $\Hc$, a set $S_c
\subseteq \nodesc(\Hc)$, and a partial role assignment $\m{ra}
\subseteq S_c \rightarrow R$, we say that the heap $\Hc$ is
{\em consistent with $\m{ra}$ for $S_c$}, and write
$\consistentWith(\m{ra},\Hc,S_c)$, iff there exists a
(total) role assignment $\rhoc : \nodesc(\Hc) \to R_0$ such
that $\m{ra}
\subseteq \rhoc$ and for every object $o \in S_c$ the
predicate $\locallyConsistent(o,\Hc,\rhoc,S_c)$ is satisfied.
\end{definition}

\subsection{Role Consistency} \label{sec:safetyCond}

We are now able to precisely state the role consistency
requirements that must be satisfied for program execution.
The role consistency requirements are in the fourth row of
Figures~\ref{fig:operationalSemantics}
and~\ref{fig:interProcOpSem}.  We assume the operational
semantics is extended with transitions leading to a program
state with heap $\errorheap$ whenever role consistency is
violated.

\subsubsection{Offstage Consistency} At
every program point, we require
$\consistent(\Hc,\offstagec(\Hc))$ to be satisfied.  This
means that offstage objects have correct roles, but onstage
objects may have their role temporarily violated.

\subsubsection{Reference Removal Consistency} \label{sec:refRemovalCond}
The Store statement \vv{x.f=y} has the following safety
precondition.  When a reference $\tu{o_x,f,o_f} \in \Hc$ for
$\tu{\m{proc}_j,\vv{x},o_x} \in \Hc$, and
$\tu{o_x,\vv{f},o_f} \in \Hc$ is removed from the heap, both
$o_x$ and $o_f$ must be referenced from the current
procedure activation record.  It is sufficient to verify
this condition for $o_f$, as $o_x$ is already onstage by
definition.  The reference removal consistency condition
enables the completion of the role change for $o_f$ after
the reference $\tu{o_x,f,o_f}$ is removed and ensures that
heap references are introduced and removed only between
onstage objects.






\subsubsection{Procedure Call Consistency} \label{sec:procCallCons}
Our programming model ensures role consistency across
procedure calls using the following protocol.  

A procedure call $\m{proc}'(x_1, ..., x_p)$ in
Figure~\ref{fig:interProcOpSem} requires the role
consistency precondition $\consistentWith(\m{ra}, H_c,
S_c)$, where the partial role assignment $\m{ra}$ requires
objects $o_k$, corresponding to parameters $x_k$, to have
roles $\prerole_k(\m{proc}')$ expected by the callee, and
$S_c =
\offstage(H_c)
\cup \{ o_k \}_k$ for $\tu{\m{proc}_j, x_k, o_k} \in \Hc$.

To ensure that the callee $\m{proc}'_j$ never observes
incorrect roles, we impose an {\em accessibility condition}
for the callee's Load statements (see the fourth column of
Figure~\ref{fig:operationalSemantics}).  The accessibility
condition prohibits access to any object $o$ referenced by
some local variable of a stack frame other than
$\m{proc}'_j$, unless $o$ is referenced by some parameter of
$\m{proc}'_j$.  Provided that this condition is not
violated, the callee $\m{proc}'_j$ only accesses objects
with correct roles, even though objects that it does not
access may have incorrect roles.  In
Section~\ref{sec:interAlg} we show how the role analysis
ensures that the accessibility condition is never violated.

At the procedure exit point
(Figure~\ref{fig:interProcOpSem}), we require correct roles
for all objects referenced by the current activation frame
$\m{proc}'_j$. This implies that heap operations performed
by $\m{proc}'_j$ preserve heap consistency for all objects
accessed by $\m{proc}'_j$.


\begin{figure*}

\begin{squeeze}
\[\begin{array}{|c|l|c|c|c|} \hline
\mbox{Statement} &
\mbox{Transition} &
\mbox{Constraints} &
\mbox{Role Consistency} \\ \hline \hline
p: \vv{roleCheck(}x_1,\ldots,x_n,\m{ra}\vv{)}                           &
\transition{\tu{p\atproc, \Hc}}
           {\tu{p'\atproc, \Hc}}                                        &
\vstack{
        \tu{p,p'} \in \CFGE
       }                                                                &
\vstack{
        \consistentWith(\m{ra},\Hc,S), \\
        \ivstack{ S & = \offstagec(\Hc) \cup {} \\
                    & \{ o \mid \tu{\nte{proc}_i,x_k,o} \in \Hc \}
                }
       }                                                                \\ \hline
\end{array}\]
\end{squeeze}
\caption{\label{fig:explicitCheckOpSem}
         {\bf Operational Semantics of Explicit Role Check}}
\end{figure*}

\subsubsection{Explicit Role Check} The programmer can specify a
stronger invariant at any program point using statement
$\vv{roleCheck}(x_1,\ldots,x_p,\m{ra})$.  As
Figure~\ref{fig:explicitCheckOpSem} indicates,
\vv{roleCheck} requires the
$\consistentWith(\m{ra},\Hc,S_c)$ predicate to be satisfied
for the supplied partial role assignment $\m{ra}$ where $S_c
= \offstage(\Hc) \cup \{o_k\}_k$ for objects $o_k$
referenced by given local variables $x_k$.

\subsection{Instrumented Semantics}
\label{sec:rolePreserving}

We expect the programmer to have a specific role assignment
in mind when writing the program, with this role assignment
changing as the statements of the program change the
referencing relationships.  So when the programmer wishes to
change the role of an object, he or she writes a program
that brings the object onstage, changes its referencing
relationships so that it plays a new role, then puts it
offstage in its new role. The roles of other objects do
not change.\footnote{An extension to the programming model
supports {\em cascading role changes} in which a single role
change propagates through the heap changing the roles of
offstage objects, see Section~\ref{sec:cascadingChange}.}

To support these programmer expectations, we introduce an
augmented programming model in which the role assignment
$\rhoc$ is conceptually part of the program's state.  The
role assignment changes only if the programmer changes it
explicitly using the \vv{setRole} statement.  The augmented
programming model has an underlying {\em instrumented
semantics} as opposed to the {\em original semantics}.


\begin{figure*}
\begin{squeeze}
\[\begin{array}{|c|l|c|c|c} \hline
\mbox{Statement} &
\mbox{Transition} &
\mbox{Constraints} &
\mbox{Role Consistency} \\ \hline \hline
p: \vv{x=new}                                                 &
\transition{\tu{p\atproc,
              \Hc \withp{\tu{\lroot,\vv{x},o_x}},\rhoc}}
           {\tu{p'\atproc,\Hc',\rhoc'}}                         &
\vstack{\vv{x} \in \proclocal(\m{proc}), \\
        o_n \mbox{ fresh}, \\
        \tu{p,p'} \in \CFGEP, \\
        \Hc' = \Hc \\
                \uplus \{ \tu{\lroot,\vv{x},o_n} \}  \\
        \uplus \{o_n\}\times F \times \{\m{null}\}, \\
        \rhoc' = \rhoc[o_n \mapsto \unknownRole]
       }                                                        &
\vstack{
        \consistentWith(\rhoc',\Hc',\offstagec(\Hc'))
       }                                                        \\ \hline
\vstack{ p : \\ \vv{setRole(x:r)}  }                                    &
\transitionc{\tu{p\atproc,\Hc,\rhoc}}
            {\tu{p'\atproc,\Hc,\rhoc'}}                                 &
\vstack{
        \vv{x} \in \proclocal(\m{proc}_i), \\
        \tu{\m{proc}_i,\vv{x},o_x} \in \Hc, \\
        \rhoc' = \rhoc[o_x \mapsto \vv{r}],\\
        \tu{p,p'} \in \CFGE
       }                                                                &
\vstack{
        \consistentWith(\rhoc',\Hc,\offstagec(\Hc))
       }                                                                \\ \hline
\vstack{p: \nte{stat}
       }                                                                &
\transitionc{\tu{s,\Hc,\rhoc}}
            {\tu{s',\Hc',\rhoc}}                                        &
\vstack{ 
        \tu{s,\Hc} \ra \tu{s',\Hc'} 
       }                                                                &
\vstack{
        P \land \consistentWith(\rhoc \cup \nte{ra},\Hc'',S) \\
        \mbox{ for every original condition } \\
        P \land \consistentWith(\nte{ra},\Hc'',S)
       }                                                                \\ \hline
\end{array}\]
\end{squeeze}

\caption{Instrumented Semantics}
\label{fig:instrumentedSemantics}
\end{figure*}

\begin{example} \label{exa:manyRoles}
The original semantics allows asserting different roles at
different program points even if the structure of the heap
was not changed, as in the following procedure \vv{foo}.
\begin{verbatim}
role A1 { fields f : B1; }
role B1 { slots A1.f; }
role A2 { fields f : B2; }
role B2 { slots A2.f; }
procedure foo()
var x, y;
{
  x = new;  y = new;
  x.f = y;
  roleCheck(x,y, x:A1,y:B1);
  roleCheck(x,y, x:A2,y:B2);
}
\end{verbatim}
Both role checks would succeed since each of the specified
partial role assignments can be extended to a valid role
assignment.  On the other hand, the check
$\vv{roleCheck(x,y, x:A1,y:B2)}$ would fail.

The procedure \vv{foo} in the instrumented semantics can be
written as folllows.
\begin{verbatim}
procedure foo()
var x, y;
{
  x = new;  y = new;
  x.f = y;
  setRole(x:A1);  setRole(y:B1);
  roleCheck(x,y, x:A1,y:B1);
  setRole(x:A2);  setRole(y:B2);
  roleCheck(x,y, x:A2,y:B2);
}
\end{verbatim}
The \vv{setRole} statement makes the role change of object
explicit.
\end{example}

The instrumented semantics extends the concrete heap $\Hc$
with a role assignment $\rhoc$.
Figure~\ref{fig:instrumentedSemantics} outlines the changes
in instrumented semantics with respect to the original
semantics.  We introduce a new statement
$\vv{setRole(x:r)}$, which modifies a role assignment
$\rhoc$, giving $\rhoc[o_x
\mapsto r]$, where $o_x$ is the object referenced by
$\vv{x}$.  All statements other than $\vv{setRole}$ preserve
the current role assignment.  For every consistency
condition $\consistentWith(\m{ra},\Hc,S_c)$ in the original
semantics, the instrumented semantics uses the corresponding
condition $\consistentWith(\rhoc \cup
\m{ra},\Hc,S_c)$ and fails if $\rhoc$ is not an extension
of $\m{ra}$.  Here we consider $\consistent(\Hc,S)$ to be a
shorthand for $\consistentWith(\emptyset,\Hc,S)$.
For example, the new role consistency condition for the Copy
statement \vv{x=y} is $\consistentWith(\rhoc, \Hc,
\offstage(\Hc))$.  The New statement assigns an identifier
$\unknownRole$ to the newly created object $o_n$.  By
definition, a node with $\unknownRole$ does not satisfy the
$\locallyConsistent$ predicate.  This means that
\vv{setRole} must be used to set a a valid role of $o_n$
before $o_n$ moves offstage.

By introducing an instrumented semantics we are not
suggesting an implementation that explicitly stores roles of
objects at run-time.  We instead use the instrumented
semantics as the basis of our role analysis and ensure that
all role checks can be statically removed.  Because the
instrumented semantics is more restrictive than the original
semantics, our role analysis is a conservative approximation
of both the instrumented semantics and the original
semantics.

\section{Intraprocedural Role Analysis} \label{sec:intraAlg}

This section presents an intraprocedural role analysis
algorithm.  The goal of the role analysis is to statically
verify the role consistency requirements described in the
previous section.

The key observation behind our analysis algorithm is that we
can incrementally verify role consistency of the concrete
heap $\Hc$ by ensuring role consistency for every node when
it goes offstage.  This allows us to represent the
statically unbounded offstage portion of the heap using
summary nodes with ``may'' references.  In contrast, we use
a ``must'' interpretation for references from and to onstage
nodes.  The exact representation of onstage nodes allows the
analysis to verify role consistency in the presence of
temporary violations of role constraints.

Our analysis representation is a graph in which nodes
represent objects and edges represent references between
objects. There are two kinds of nodes: {\em onstage nodes}
represent onstage objects, with each onstage node
representing one onstage object; and {\em offstage nodes},
with each offstage node corresponding to a set of objects
that play that role. To increase the precision of the
analysis, the algorithm occasionally generates multiple
offstage nodes that represent disjoint sets of objects
playing the same role.  Distinct offstage objects with the
same role $r$ represent disjoint sets of objects of role $r$
with different reachability properties from onstage nodes.

We frame role analysis as a data-flow analysis operating on
a distributive lattice $\mathcal{P}(\RoleGraphs)$ of sets of
role graphs with set union $\cup$ as the join operator.  In
this section we present an algorithm for intraprocedural
analysis.  We use $\concproc$ to denote the topmost
activation record in a concrete heap $\Hc$.  In
Section~\ref{sec:interAlg} we generalize the algorithm
to the compositional interprocedural analysis.

\subsection{Abstraction Relation} \label{sec:anaAbstraction}

Every data-flow fact $\GS \subseteq \RoleGraphs$ is a set of
role graphs $G \in \GS$.  Every role graph $G \in
\RoleGraphs$ is either a bottom role graph $\Gb$
representing the set of all concrete heaps (including
$\errorheap$), or a tuple $G = \tu{H,\rho,K}$ representing
non-error concrete heaps, where
\begin{itemize} \itemsep=-0em
\item $H \subseteq N \times F \times N$ is the
      abstract heap with nodes $N$ representing objects
      and fields $F$.
The abstract heap $H$ represents heap references
      $\tu{n_1,f,n_2}$ and variables of the currently
      analyzed procedure $\tu{\abstproc,x,n}$ where $x \in
      \proclocal(\abstproc)$.  Null references are
      represented as references to abstract node
      $\nullAbstNode$.  
      We define abstract onstage nodes $\onstage(H) = \{
      n \mid \tu{\abstproc,x,n} \in H, x \in
      \proclocal(\abstproc) \cup \procparam(\abstproc) \}$ and
      abstract offstage nodes $\offstage(H) = \nodes(H)
      \setminus \onstage(H) \setminus \{\abstproc, \nullAbstNode\}$.
\item $\rho : \nodes(H) \to R_0$ is an abstract role assignment,
      $\rho(\nullAbstNode) = \nullRole$;
\item $K : \nodes(H) \to \{i, s\}$ indicates the kind of each node; 
      when $K(n) = i$, then $n$ is an individual node
      representing at most one object, and when $K(n) = s$,
      $n$ is a summary node representing zero or more
      objects.  We require $K(\abstproc) = K(\nullAbstNode)
      = i$, and require all onstage nodes to be individual,
      $K[\onstage(H)] \subseteq \{i\}$.
\end{itemize}
The abstraction relation $\abstrel$ relates a pair
$\tu{\Hc,\rhoc}$ of concrete heap and concrete role
assignment with an abstract role graph $G$.
\begin{definition} \label{def:abstrel}
We say that an abstract role graph $G$ represents concrete
heap $\Hc$ with role assignment $\rhoc$ and write
$\tu{\Hc,\rhoc} \abstrel G$, iff $G = \Gb$ or: 
$\Hc \neq \errorheap$, 
$G = \tu{H,\rho,K}$, 
and there exists a
function $\homo : \nodesc(\Hc) \to \nodes(H)$ such that
\begin{enumerate} \itemsep=0em
\item[1)] $\Hc$ is role consistent:
      $\consistentWith(\rhoc,\Hc,\offstage(\Hc))$,
\item[2)] identity constraints of onstage nodes with offstage nodes
      hold: if $\tu{o_1,f,o_2} \in \Hc$ and $\tu{o_2,g,o_3}
      \in \Hc$ for $o_1 \in \onstage(\Hc)$,
      $o_2 \in \offstage(\Hc)$, and \newline
      $\tu{f,g} \in \identities(\rhoc(o_1))$,
      then $o_3 = o_1$;
\item[3)] $\homo$ is a graph homomorphism:
      if $\tu{o_1,f,o_2} \in \Hc$ then
      $\tu{\homo(o_1),f,\homo(o_2)} \in H$;
\item[4)] an individual node represents at most one concrete object:
      $K(n) = i$ implies $|\homo^{-1}(n)| \leq 1$;
\item[5)] $\homo$ is bijection on edges which originate or
      terminate at onstage nodes: if $\tu{n_1,f,n_2} \in H$
      and $n_1 \in \onstage(H)$ or $n_2 \in \onstage(H)$,
      then there exists exactly one \newline $\tu{o_1,f,o_2} \in \Hc$
      such that $\homo(o_1) = n_1$ and $\homo(o_2) = n_2$;
\item[6)] $\homo(\nullObj) = \nullAbstNode$ and $\homo(\concproc) = \abstproc$;
\item[7)] the abstract role assignment $\rho$ corresponds to the
      concrete role assignment: 
      $\rhoc(o) = \rho(\homo(o))$ for every object $o \in
      \nodesc(\Hc)$.
\end{enumerate}
\end{definition}
Note that the error heap $\errorheap$ can be represented
only by the bottom role graph $\Gb$.  The analysis uses
$\Gb$ to indicate a potential role error.

Condition 3) implies that role graph edges are a
conservative approximation of concrete heap references.
These edges are in general ``may'' edges.  Hence it is
possible for an offstage node $n$ that $\tu{n,f,n_1}$,
$\tu{n,f,n_2} \in H$ for $n_1
\neq n_2$.  This cannot happen when $n \in \onstage(H)$
because of 5).  Another consequence of 5) is that an edge in
$H$ from an onstage node $n_0$ to a summary node $n_s$
implies that $n_s$ represents at least one object.
Condition 2) strengthens 1) by requiring certain identity
constraints for onstage nodes to hold, as explained in
Section~\ref{sec:nodeCheck}.

\begin{figure}
\begin{center}
\epsfig{file=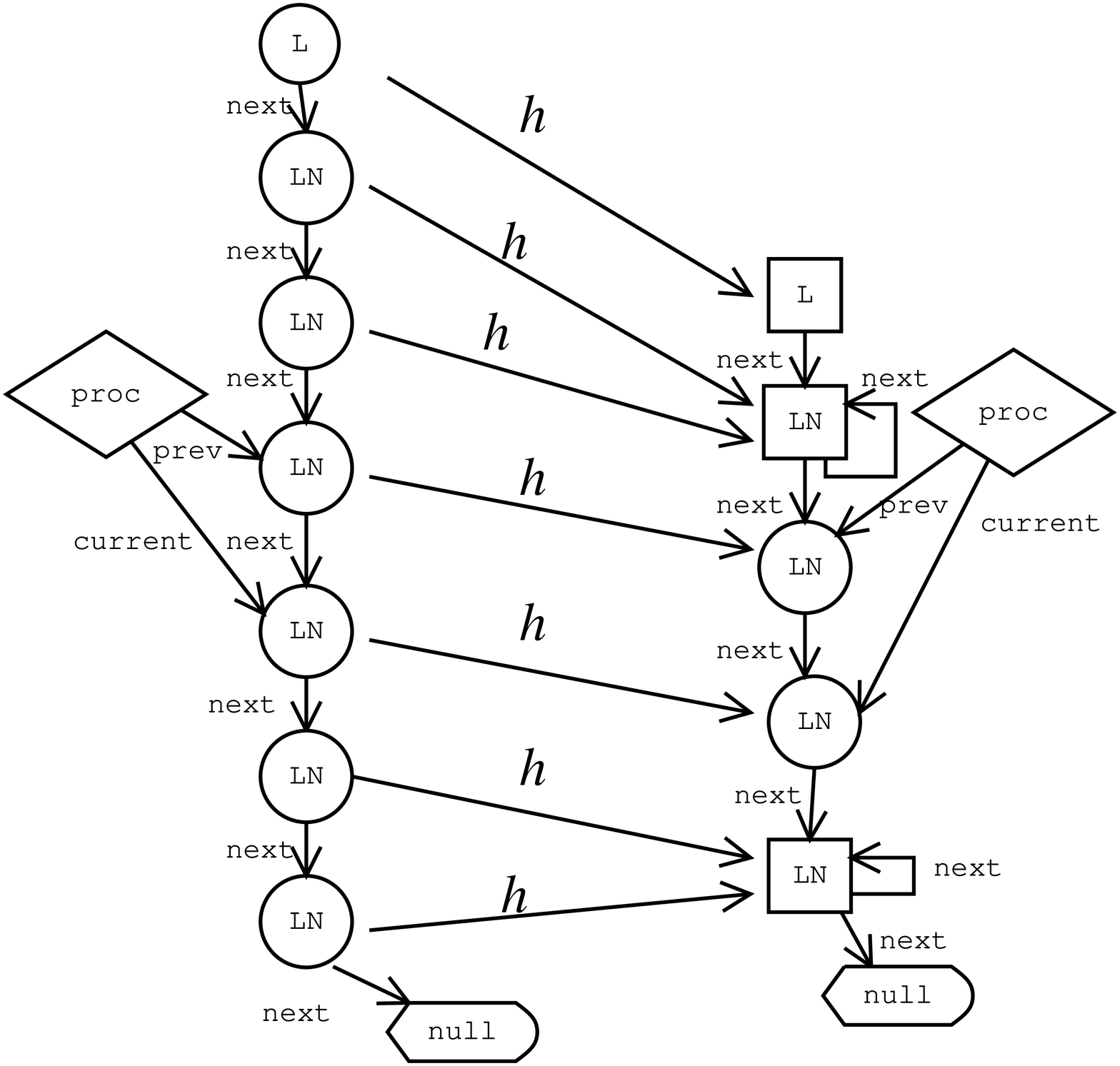, width=3.5in, height=4in}
\end{center}
\caption{Abstraction Relation}
\label{fig:abstRel}
\end{figure}

\begin{example}
Consider the following role declaration for an acyclic list.
\begin{verbatim}
role L { // List header
  fields first : LN | null;
}
role LN { // List node
  fields next : LN | null;
  slots LN.next | L.first;
  acyclic next;
}
\end{verbatim}
Figure~\ref{fig:abstRel} shows a role graph and one of the
concrete heaps represented by the role graph via
homomorphism $h$.  There are two local variables, \vv{prev}
and \vv{current}, referencing distinct onstage objects.
Onstage objects are isomorphic to onstage nodes in the role
graph.  In contrast, there are two objects mapped to each of
the summary nodes with role \vv{LN} (shown as
\vv{LN}-labelled rectangles in Figure~\ref{fig:abstRel}).  Note
that the sets of objects mapped to these two summary nodes are
disjoint.  The first summary
\vv{LN}-node represents objects stored in the list before
the object referenced by \vv{prev}.  The second summary
\vv{LN}-node represents objects stored in the list after the
object referenced by
\vv{current}.
\end{example}

\subsection{Transfer Functions}

The key complication in developing the transfer functions
for the role analysis is to accurately model the movement of
objects onstage and offstage.  For example, a load
statement \vv{x=y.f} may cause the object referred to by
\vv{y.f} to move onstage.  In addition, if \vv{x} was the 
only reference to an onstage object $o$ before the statement
executed, object $o$ moves offstage after the execution of
the load statement, and thus must satisfy the
$\locallyConsistent$ predicate.

The analysis uses an expansion relation $\expandeqwith{}$ to
model the movement of objects onstage and a contraction
relation $\contracteqwith{}$ to model the movement of
objects offstage.  The expansion relation uses the invariant
that offstage nodes have correct roles to generate possible
aliasing relationships for the node being pulled onstage.
The contraction relation establishes the role invariants for
the node going offstage, allowing the node to be merged into
the other offstage nodes and represented more compactly.

We present our role analysis as an abstract execution
relation $\transrel{st}$.  The abstract execution ensures
that the abstraction relation $\alpha$ is a forward
simulation relation \cite{LynchVaandrager95Forward} from the
space of concrete heaps with role assignments to the set
$\RoleGraphs$.  The simulation relation implies that the
traces of $\transrel{}$ include the traces of the
instrumented semantics $\rac$.  To ensure that the program
does not violate constraints associated with roles, it is
thus sufficient to guarantee that $\Gb$ is not reachable via
$\transrel{}$.

\begin{figure}
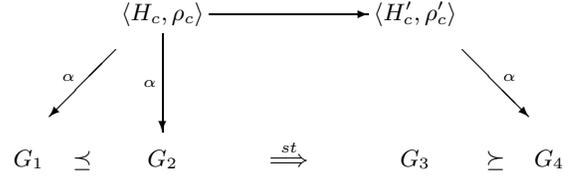

\input arrow.tex
$$\sarrowlength=.42\harrowlength
\commdiag{
&&\tu{H_c, \rho_c}&\mapright&\tu{H_c', \rho_c'}\cr
&\arrow(-1,-1)\lft{\alpha}&\mapdown\lft{\alpha}&&&\arrow(1,-1)\rt{\alpha}\cr
G_1&\expandeqwith{}&G_2&
\mathop{\Longrightarrow}\limits^{st}
&G_3&\contracteqwith{}&G_4\cr
}$$
\caption{Simulation Relation Between Abstract and Concrete Execution}
\label{fig:simRel}
\end{figure}

To prove that $\Gb$ is not reachable in the abstract execution,
the analysis computes for every program point $p$ a set of
role graphs $\GS$ that conservatively approximates the
possible program states at point $p$.  The transfer function
for a statement $\m{st}$ is an image
$\tr{\m{st}}(\GS) = \{ G'
\mid G \in \GS, G \transrel{st} G' \} $.  The analysis
computes the relation
$\transrel{st}$ in three steps:
\begin{enumerate}
\item ensure that the relevant nodes are instantiated using expansion relation
$\expandeqwith{}$ (Section~\ref{sec:expansion});
\item perform symbolic execution
$\symexec{st}$ of the statement $\vv{st}$ (Section~\ref{sec:symExec});
\item merge nodes if
needed using contraction relation $\contracteq{}$ to keep
the role graph bounded (Section~\ref{sec:contraction}).
\end{enumerate}
Figure~\ref{fig:simRel} shows how the abstraction relation
$\alpha$ relates $\expandeq$, $\symexec{st}$, and
$\contracteqwith{}$ with the concrete execution $\rac$ in
instrumented semantics.  Assume that a concrete heap
$\tu{\Hc,\rhoc}$ is represented by the role graph $G_1$.
Then one of the role graphs $G_2$ obtained after expansion
remains an abstraction of $\tu{\Hc,\rhoc}$.  The symbolic
execution $\symexec{st}$ followed by the contraction
relation $\contracteqwith{}$ corresponds to the instrumented
operational semantics $\rightarrow$.

\begin{figure*}
\[\begin{array}{|c|c|c|} \hline 
\mbox{Transition}
&
\mbox{Definition}
&
\mbox{Conditions}
\\ \hline
\tu{H,\rho,K} \transrel{x=y.f}
G'
&
\tu{H,\rho,K} \expandeqwith{n_y,f}
G_1           \symexec{x=y.f}
G_2           \contracteqwith{n_x}
G'
&
\tu{\abstproc,\vv{x},n_x}, \tu{\abstproc,\vv{y},n_y} \in H
\\ \hline
\tu{H,\rho,K} \transrel{x=y}
G'
&
\tu{H,\rho,K} \symexec{x=y}
G_1           \contracteqwith{n_1}
G'
&
\tu{\abstproc,\vv{x},n_1} \in H
\\ \hline
\tu{H,\rho,K} \transrel{x=new}
G'
&
\tu{H,\rho,K} \symexec{x=new}
G_1           \contracteqwith{n_1}
G'
&
\tu{\abstproc,\vv{x},n_1} \in H
\\ \hline
\tu{H,\rho,K} \transrel{s} G'
&
\tu{H,\rho,K} \symexec{s}
G'
&
\vstack{
        s \in \{ \vv{x.f=y},\\
            \vv{test(c)},\\
            \vv{setRole(x:r)},\\
            \vv{roleCheck($x_{1..p}, \m{ra}$)} \} \\
       }
\\ \hline
\end{array}\]

\caption{Abstract Execution $\transrel{}$}
\label{fig:transrel}
\end{figure*}

Figure~\ref{fig:transrel} shows rules for the abstract
execution relation $\transrel{st}$.  Only Load statement
uses the expansion relation, because the other statements
operate on objects that are already onstage.  Load, Copy,
and New statements may remove a local variable reference
from an object, so they use contraction relation to move the
object offstage if needed.  For the rest of the statements,
the abstract execution reduces to symbolic execution $\symexec{}$
described in Section~\ref{sec:symExec}.

\paragraph{Nondeterminism and Failure}
The $\transrel{st}$ relation is not a function because the
expansion relation $\expandeq$ can generate a set of role
graphs from a single role graph.  Also, there might be no
$\transrel{st}$ transitions originating from a given state
$G$ if the symbolic execution $\symexec{}$ produces no
results.  This corresponds to a trace which cannot be
extended further due to a \vv{test} statement which fails in
state $G$.  This is in contrast to a transition from $G$ to
$\Gb$ which indicates a potential role consistency violation
or a null pointer dereference.  We assume that $\symexec{}$
and $\contracteq$ relations contain the transition
$\tu{\Gb,\Gb}$ to propagate the error role graph.  In most
cases we do not write the explicit transitions to error
states.

\subsubsection{Expansion} \label{sec:expansion}


\begin{figure*}
\[\begin{array}{|c|c|c|} \hline
\mbox{Transition}
&
\mbox{Definition}
&
\mbox{Condition}
\\ \hline \hline
\tu{H,\rho,K} \expandeqwith{n,f}
\tu{H,\rho,K}
&
&
\tu{n,f,n'} \in H,  n' \in \onstage(H)
\\ \hline
\tu{H,\rho,K} \expandeqwith{n,f}
G'
&
\tu{H,\rho,K}         \instantiate{n'}{n_0}
\tu{H_1,\rho_1,K_1}   \splt{n_0}
G'
&
\vstack{
   \tu{n,f,n'} \in H, n' \in \offstage(H) \\
   \tu{n,f,n_0} \in H_1 \\
}
\\ \hline
\end{array}\]
\caption{Expansion Relation}
\label{fig:expansionRelation}
\end{figure*}

Figure~\ref{fig:expansionRelation} shows the expansion
relation $\expandeqwith{n,f}$.  Given a role graph
$\tu{H,\rho,K}$ expansion attempts to produce a set of role
graphs $\tu{H',\rho',K'}$ in each of which $\tu{n,f,n_0} \in
H'$ and $K(n_0) = i$.  Expansion is used in abstract
execution of the Load statement.  It first checks for null
pointer dereference and reports an error if the check fails.
If $\tu{n,f,n'} \in H$ and $K(n') = i$ already hold, the
expansion returns the original state.  Otherwise,
$\tu{n,f,n'} \in H$ with $K(n') = s$.  In that case, the
summary node $n'$ is first instantiated using instantiation
relation $\instantiate{n'}{n_0}$.  Next, the split relation
$\splt{n_0}$ is applied.  Let $\rho(n_0) = r$.  The split
relation ensures that $n_0$ is not a member of any cycle of
offstage nodes which contains only edges in $\acyclic(r)$.
We explain instantiation and split in more detail below.

\begin{figure*}
\[\begin{array}{|c|l|} \hline
\tu{H,\rho,K} \instantiate{n'}{n_0} \tu{H',\rho',K'} 
& \begin{array}{l} 
H' = H \setminus H_0 \cup H'_0 \cup H'_1 \\
n' \notin \nodes(H'), \mbox{ if } K(n') = i \\
\rho' = \rho[n_0 \mapsto \rho(n')] \\
K' = K[n_0 \mapsto i] \\
\localCheck(n_0,\tu{H',\rho',K'}) \\
H_0 \subseteq H \cap \big( \onstage(H) \times F \times \{n'\} \cup
                              \{n'\} \times F \times \onstage(H) \big) \\
H_1 \subseteq H \cap \big( \offstage(H) \times F \times \{n'\} \cup
                              \{n'\} \times F \times \offstage(H) \big) \\
H'_0 = \swing(n',n_0,H_0) \\
H'_1 \subseteq \swing(n',n_0,H_1) \\
\end{array} \\ \hline
\end{array}\]
\[\begin{array}{rcl}
\swing(n_{\m{old}},n_{\m{new}},H) & = & \{ \tu{n_{\m{new}},f,n} \mid \tu{n_{\m{old}},f,n} \in H \} \cup 
                       \{ \tu{n,f,n_{\m{new}}} \mid \tu{n,f,n_{\m{old}}} \in H \} \cup {} \\
                  &  & \{ \tu{n_{\m{new}},f,n_{\m{new}}} \mid \tu{n_{\m{old}},f,n_{\m{old}}} \in H \} \\
\end{array}\]
\caption{Instantiation Relation $\instantiate{}{}$
\label{fig:instantiation}}
\end{figure*}

\paragraph{Instantiation}
Figure~\ref{fig:instantiation} presents the instantiation
relation.  Given a role graph $G = \tu{H,\rho,K}$,
instantiation $\instantiate{n'}{n_0}$ generates the set of
role graphs $\tu{H',\rho',K'}$ such that each concrete heap
represented by $\tu{H,\rho,K}$ is represented by one of the
graphs $\tu{H',\rho',K'}$.  Each of the new role graphs
contains a fresh individual node $n_0$ that satisfies
$\localCheck$.  The edges of $n_0$ are a subset of edges
from and to $n'$.

Let $H_0$ be a subset of the references between $n'$ and
onstage nodes, and let $H_1$ be a subset of the references
between $n'$ and offstage nodes.  References in $H_0$ are
moved from $n'$ to the new node $n_0$, because they
represent at most one reference, while references in $H_1$
are copied to $n_0$ because they may represent multiple
concrete heap references.  Moving a reference is formalized
via the $\swing$ operation in
Figure~\ref{fig:instantiation}.

The instantiation of a single graph can generate multiple
role graphs depending on the choice of $H'_0$ and $H'_1$.
The number of graphs generated is limited by the existing
references of node $n'$ and by the $\localCheck$ requirement
for $n_0$.  This is where our role analysis takes advantage
of constraints associated with role definitions to reduce
the number of aliasing possibilities that need to be
considered.

\begin{figure*}
\[\begin{array}{ll}
     \tu{H,\rho,K} \splt{n_0} \tu{H,\rho,K},
 & \acycCheck(n_0,\tu{H,\rho,K},\offstage(H)) \\
     \tu{H,\rho,K} \splt{n_0} \tu{H',\rho',K'}, 
 & \lnot \acycCheck(n_0,\tu{H,\rho,K},\offstage(H))
\end{array}\]
\[\begin{array}{rl}
\mbox{where}
\\
& H' = (H \setminus \Hcyc) 
     \cup \Hoff
     \cup \BfNR \cup \BfR 
     \cup \BtNR \cup \BtR
     \cup \Nf \cup \Nt
\\
& \Hcyc = \{ \tu{n_1,f,n_2} \mid n_1 \mbox{ or } n_2 \in \Scyc \}
\\
& 
\begin{array}{@{}rcccl}
  \Hoff & = & \large\{ \> \tu{n'_1,f,n'_2} & \mid &  n_1 = c(n'_1), n_2 = c(n'_2), \\
        &   &                     &      &  n_1, n_2 \in \offstage_1(H), n_1 \mbox{ or } n_2 \in \Scyc, \\
        &   &                     &      &  \tu{n_1,f,n_2} \in H \large\>\}  \\
        &   & \multicolumn{3}{l}{\setminus (\SR \times \acyclic(r) \times \SNR)} \\
\end{array}
\\
& H \cap (\onstage(H) \times F \cup \{n_0\} \times \acyclic(r)) \times \Scyc = \AfNR \uplus \AfR
\\
& H \cap \Scyc \times (\acyclic(r) \times \{n_0\} \cup F \times \onstage(H))  = \AtNR \uplus \AtR
\\
& \BfNR = \{ \tu{n_1,f,\hNR(n_2)} \mid \tu{n_1,f,n_2} \in \AfNR \}
\\
& \BfR  = \{ \tu{n_1,f,\hR(n_2)} \mid \tu{n_1,f,n_2} \in \AfR \}
\\
& \BtNR = \{ \tu{\hNR(n_1),f,n_2} \mid \tu{n_1,f,n_2} \in \AtNR \}
\\
& \BtR = \{ \tu{\hR(n_1),f,n_2} \mid \tu{n_1,f,n_2} \in \AtR \}
\\
& \Nf = \{ \tu{n_0,f,n'} \mid n' \in \SR, \tu{n_0,f,c(n')} \in H, f \in \acyclic(r) \}
\\
& \Nt = \{ \tu{n',f,n_0} \mid n' \in \SNR, \tu{c(n'),f,n_0} \in H, f \in \acyclic(r) \}
\\
& \Scyc = \{ n \mid \exists n_1,\ldots,n_{p-1} \in \offstage(H) : \\
&   \qquad \tu{n_0,f_0,n_1},\ldots,\tu{n_k, f_k, n}, \tu{n, f_{k+1}, n_{k+2}}, \tu{n_{p-1},f_{p-1},n_0} \in H, \\
&   \qquad f_0,\ldots,f_{p-1} \in \acyclic(r) \} 
\\
& \offstage_1(H) = \offstage(H) \setminus \{n_0\}
\\
& r = \rho(n_0)
\\
\\
& \rho'(c(n)) = \rho(n)
\\
& K'(c(n)) = K(n)
\\
\end{array}\]
\caption{Split Relation}
\label{fig:splitFormula}
\end{figure*}

\paragraph{Split}
The split relation is important for verifying operations on
data structures such as skip lists and sparse matrices.  It
is also useful for improving the precision of the initial
set of role graphs on procedure entry
(Section~\ref{sec:roleGraphsOnEntry}).

The goal of the split relation is to exploit the acyclicity
constraints associated with role definitions.  After a node
$n_0$ is brought onstage, split represents the acyclicity
condition of $\rho(n_0)$ explicitly by eliminating
impossible paths in the role graph.  It uses additional
offstage nodes to encode the reachability information
implied by the acyclicity conditions.  This information can
then be used even after the role of node $n_0$ changes.  In
particular, it allows the acyclicity condition of $n_0$ to
be verified when $n_0$ moves offstage.

\begin{figure}
\begin{center}
\epsfig{file=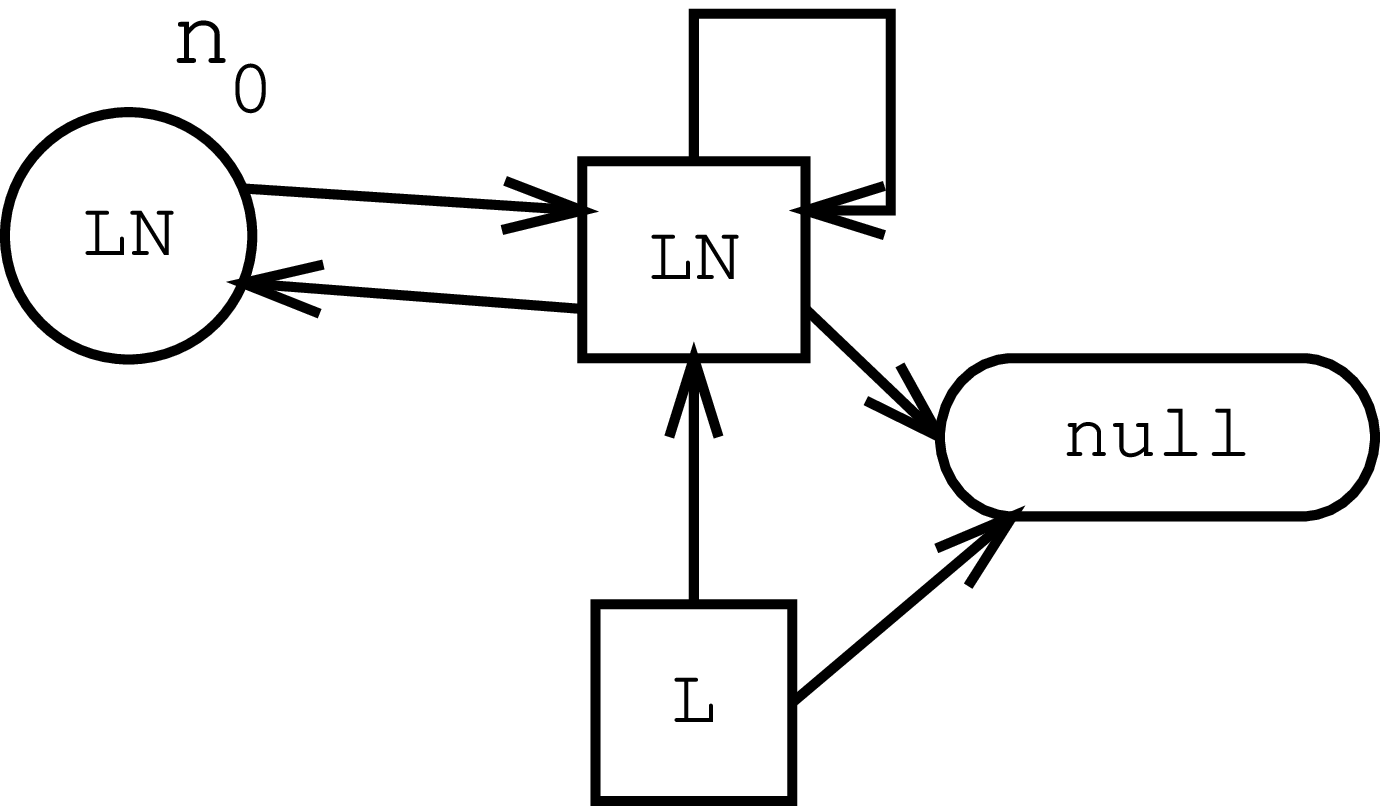,width=1.8in}
\vspace{1em}
\centerline{a) Before Split}
\vspace{2em}
\epsfig{file=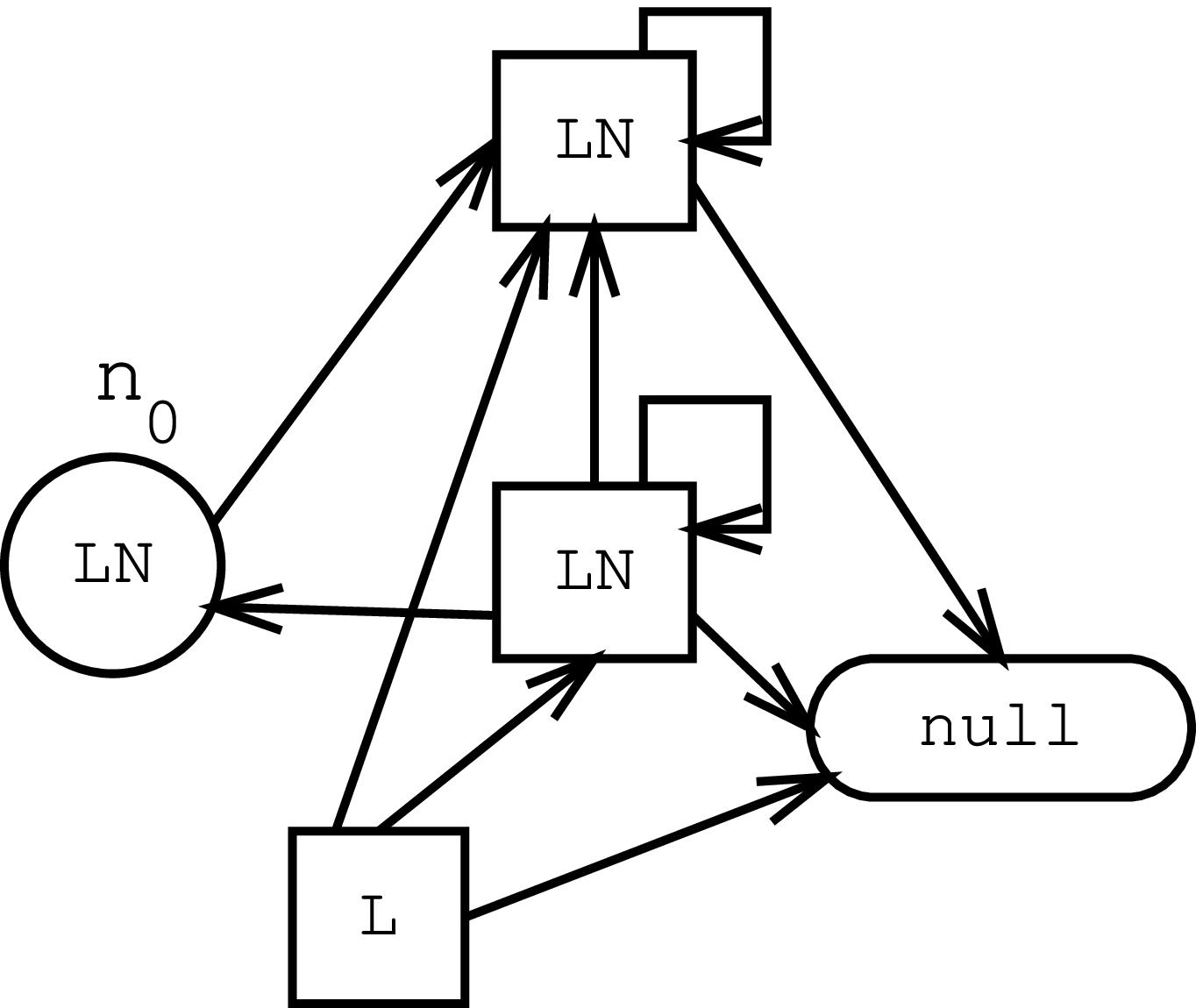,width=1.8in}
\vspace{1em}
\centerline{b) After Split}
\vspace{2em}
\caption{A Role Graph for an Acyclic List}
\label{fig:split}
\end{center}
\end{figure}

\begin{example}
Consider a role graph for an acyclic list with nodes \vv{LN}
and a header node \vv{L}.  The instantiated node $n_0$ is in
the middle of the list.  Figure~\ref{fig:split} a) shows a
role graph with a single summary node representing all
offstage \vv{LN}-nodes.  Figure~\ref{fig:split} b) shows the
role graph after applying the split relation.  The resulting
role graph contains two \vv{LN} summary nodes.  The first
\vv{LN} summary node represents objects definitely 
reachable from $n_0$ along \vv{next} edges; the second
summary \vv{NL} node represents objects definitely not
reachable from $n_0$.
\end{example}

Figure~\ref{fig:splitFormula} shows the definition of the
split operation on node $n_0$, denoted by $\splt{n_0}$.  Let
$G =
\tu{H,\rho,K}$ be the initial role graph and $\rho(n_0) = r$.  If
$\acyclic(r) = \emptyset$, then the split operation returns
the original graph $G$; otherwise it proceeds as follows.
Call a path in graph $H$ {\em cycle-inducing} if all of its
nodes are offstage and all of its edges are in
$\acyclic(r)$.  Let $\Scyc$ be the set of nodes $n$ such
that there is a cycle-inducing path from $n_0$ to $n$ and a
cycle-inducing path from $n$ to $n_0$.

The goal of the split operation is to split the set $\Scyc$
into a fresh set of nodes $\SNR$ representing objects
definitely not reachable from $n_0$ along edges in
$\acyclic(r)$ and a fresh set of nodes $\SR$ representing
objects definitely reachable from $n_0$.  Each of the newly
generated graphs $H'$ has the following properties:
\begin{enumerate} \itemsep=0em
\item[1)] merging the corresponding nodes from $\SNR$ and $\SR$ 
      in $H'$ yields the original graph $H$;
\item[2)] $n_0$ is not a member of any cycle in $H'$ consisting of 
      offstage nodes and edges in $\acyclic(r)$;
\item[3)] onstage nodes in $H'$ have the same number of fields
      and aliases as in $H$.
\end{enumerate}
Let $S_0 = \nodes(H)
\setminus \Scyc$ and let $\hNR : \Scyc \to \SNR$ and $\hR :
\Scyc \to \SR$ be bijections.
Define a function $c : \nodes(H') \to \nodes(H)$ as follows:
\[
   c(n) = \left\{\begin{array}{rl}
          n,            & n \in S_0 \\
          \hR^{-1}(n),  & n \in \SR \\
          \hNR^{-1}(n), & n \in \SNR \\
          \end{array} \right.
\]
Then $H' \subseteq \{ \tu{n'_1,f,n'_2} \mid \tu{c(n'_1),f,c(n'_2)} \in H\}$.

Because there are two copies of $S_0$ in $H'$, there might
be multiple edges $\tu{n'_1,f,n'_2}$ in $H'$ corresponding
to an edge $\tu{c(n_1),f,c(n_2)} \in H$.

If both $n'_1$ and $n'_2$ are offstage nodes other than
$n_0$, we always include $\tu{n'_1,f,n'_2}$ in $H'$ unless
$\tu{n'_1,f,n'_2} \in \SR \times \acyclic(r) \times \SNR$.
The last restriction prevents cycles in $H'$.

For an edge $\tu{n_1,f,n_2} \in H$ where $n_1 \in
\onstage(H)$ and $n_2 \in \Scyc$ we include in $H'$ either
the edge $\tu{n_1,f,\hNR(n_2)}$ or $\tu{n_1,f,\hR(n_2)}$ but
not both.  Split generates multiple graphs $H'$ to cover
both cases.  We proceed analogously if $n_2 \in \onstage(H)$
and $n_1 \in \Scyc$.  The node $n_0$ itself is treated in
the same way as onstage nodes for $f \notin \acyclic(r)$.
If $f \in \acyclic(r)$ then we choose references {\em to}
$n_0$ to have a source in $\SNR$, whereas the reference {\em
from} $n_0$ have the target in $\SR$.

Details of the split construction are given in
Figure~\ref{fig:splitFormula}.  The intuitive meaning of the
sets of edges is the following:

\begin{tabular}{c@{$\>$}c@{$\>$}l}
$\Hoff$ & : & edges between offstage nodes \\
$\BfNR$ & : & edges from onstage nodes to $\SNR$ \\
$\BfR$  & : & edges from onstage nodes to $\SR$ \\
$\BtNR$ & : & edges from $\SNR$ to onstage nodes \\
$\BtR$  & : & edges from $\SR$ to onstage nodes \\
$\Nf$   & : & $\acyclic(r)$-edges from $n_0$ to $\SR$ \\
$\Nt$   & : & $\acyclic(r)$-edges from $\SNR$ to $n_0$ \\
\end{tabular}

\noindent
The sets $\BfNR$ and $\BfR$ are created as images
of the sets $\AfNR$ and $\AfR$ which partition edges
from onstage nodes to nodes in $\Scyc$.
Similarly, the sets $\BtNR$ and $\BtR$ are created as images
of the sets $\AtNR$ and $\AtR$ which partition edges
from nodes in $\Scyc$ to onstage nodes.

We note that if in the split operation $\Scyc = \emptyset$
then the operation has no effect and need not be performed.
In Figure~\ref{fig:split}, after performing a single split,
there is no need to split for subsequent elements of the
list.  Examples like this indicate that split will not be
invoked frequently during the analysis.



\begin{figure*}
\[\begin{array}{|c|c|} \hline
\tu{H,\rho,K} \contracteqwith{n} \tu{H,\rho,K} 
&
\vstack { 
          \exists \vv{x} \in \procvar(\m{proc}) : {} \\
          \tu{\abstproc,\vv{x},n} \in H
        }
\\ \hline
\tu{H,\rho,K} \contracteqwith{n} \normalize(\tu{H,\rho,K})
&
\nodeCheck(n,\tu{H,\rho,K},\offstage(H))
\\ \hline
\end{array}\]
\caption{Contraction Relation}
\label{fig:contraction}
\end{figure*}

\subsubsection{Contraction} \label{sec:contraction}

Figure~\ref{fig:contraction} shows the non-error transitions
of the contraction relation $\contracteqwith{n}$.  The
analysis uses contraction when a local variable reference to
node $n$ is removed.  If there are other local references to $n$,
the result is the original graph.  Otherwise $n$ has just
gone offstage, so analysis invokes $\nodeCheck$.  If the
check fails, the result is $\Gb$.  If the role check
succeeds, the contraction invokes normalization operation to
ensure that the role graph remains bounded.  For simplicity,
we use normalization whenever $\nodeCheck$ succeeds,
although it is sufficient to perform normalization only at
program points adjacent to back edges of the control-flow
graph.

\begin{figure*}
\[
   \normalize(\tu{H,\rho,K}) = \tu{H',\rho',K'}
\]
\[\begin{array}{rl}
\mbox{where}
& H' = \{ \tu{\cl{n_1},f,\cl{n_2}} \mid \tu{n_1,f,n_2} \in H \} \\
& \rho'(\cl{n}) = \rho(n) \\
& K'(\cl{n}) = \left\{\begin{array}{rl}
               i, & \cl{n} = \{n\}, K(n) = i \\
               s, & \mbox{otherwise} \\
               \end{array}\right. \\
& n_1 \sim n_2 \mbox{ iff } n_1 = n_2 \mbox{ or } \\
& \qquad \qquad (n_1,n_2 \in \offstage(H), \rho(n_1) = \rho(n_2), \\
& \qquad \qquad  \forall n_0 \in \onstage(H) : 
      (\reach(n_0,n_1) \mbox{ iff } \reach(n_0,n_2) ) \\
& \reach(n_0,n) \mbox{ iff } \exists n_1,\ldots,n_{p-1} \in \offstage(n),
                             \exists f_1,\ldots,f_p \in \acyclic(\rho(n_0)) : \\
& \qquad \qquad \tu{n_0,f_1,n_1},\ldots,\tu{n_{p-1},f_p,n} \in H \\
\end{array}\] 
\caption{Normalization}
\label{fig:normalization}
\end{figure*}

\paragraph{Normalization} 

Figure~\ref{fig:normalization} shows the normalization
relation.  Normalization accepts a role graph
$\tu{H,\rho,K}$ and produces a normalized role graph
$\tu{H',\rho',K'}$ which is a factor graph of
$\tu{H,\rho,K}$ under the equivalence relation $\sim$.  Two
offstage nodes are equivalent under $\sim$ if they have the
same role and the same reachability from onstage nodes.
Here we consider node $n$ to be reachable from an onstage node $n_0$
iff there is some path from $n_0$ to $n$ whose edges belong
to $\acyclic(\rho(n_0))$ and whose nodes are all in
$\offstage(H)$.  Note that, by construction, normalization
avoids merging nodes which were previously generated in the
split operation $\splt{}$, while still ensuring a bound on
the size of the role graph.  
For a procedure with $l$ local
variables, $f$ fields and $r$ roles the number of nodes in a
role graph is on the order of $r 2^l$ so the maximum size of
a chain in the lattice is of the order of $2^{r2^l}$.  
To
ensure termination we consider role graphs equal up to
isomorphism.  Isomorphism checking can be done efficiently
if normalization assigns canonical names to the equivalence
classes it creates.

\begin{figure*}
\newcommand{\transi}[2]{{#1} \symexec{s} {#2}}
\begin{squeeze}
\[\begin{array}{|c|c|c|} \hline 
\mbox{Statement } $s$
&
\mbox{Transition}
&
\mbox{Conditions}
\\ \hline \hline
\vv{x = y.f}
&
\transi{\tu{H \withp{\abstproc,\vv{x},n_x},\rho,K}}
       {\tu{H \withp{\abstproc,\vv{x},n_f},\rho,K}}
&
\tu{\abstproc,\vv{y},n_y}, \tu{n_y,f,n_f} \in H
\\ \hline
\vv{x.f = y}
&
\transi{\tu{H \withp{n_x,f,n_f},\rho,K}}
       {\tu{H \withp{n_x,f,n_y},\rho,K}}
&
\vstack{
        \tu{\abstproc,\vv{x},n_x},\tu{\abstproc,\vv{y},n_y} \in H \\
        n_f \in \onstage(H)
       }        
\\ \hline
\vv{x = y}
&
\transi{\tu{H \withp{\abstproc,\vv{x},n_x},\rho,K}}
       {\tu{H \withp{\abstproc,\vv{x},n_y},\rho,K}}
&
\tu{\abstproc,\vv{y},n_y} \in H
\\ \hline
\vv{x = new}
&
\transi{\tu{H \withp{\abstproc,\vv{x},n_x},\rho,K}}
       {\tu{H \withp{\abstproc,\vv{x},n_n},\rho',K}}
&
\vstack{
         n_n \mbox{ fresh} \\
         \rho' = \rho[n_n \mapsto \unknownRole] \\
       }
\\ \hline
\vv{test(c)}
&
\transi{\tu{H,\rho,K}}
       {\tu{H,\rho,K}}
&
\abstsatisfied(\vv{c},H)
\\ \hline
\vv{setRole(x:r)}
&
\transi{\tu{H,\rho,K}}
       {\tu{H,\rho[n_x \mapsto \vv{r}],K}}
&
\vstack{
  \tu{\abstproc,\vv{x},n_x} \in H \\
  \rchange(n_x,\vv{r},\tu{H,\rho,K})
}
\\ \hline
\vv{roleCheck($x_{1..p}, \m{ra}$)}
&
\transi{\tu{H,\rho,K}}
       {\tu{H,\rho,K}}
&
\vstack{
        \forall i \> \tu{\abstproc,\vv{x}_i,n_i} \in H \\
         \nodeCheck(n_i,\tu{H,\rho,K},S) \\
         S = \offstage(H) \cup \{n_i\}_i \\
         \rho(n_i) = \m{ra}(n_i) \\        
       }
\\ \hline
\end{array}\]
\[  \abstsatisfied(\vv{x==y}, \Hc) \mbox{ iff } 
       \{ o \mid \tu{\abstproc,\vv{x},o} \in \Hc \} =
       \{ o \mid \tu{\abstproc,\vv{y},o} \in \Hc \}
\]
\[  \abstsatisfied(\vv{!(x==y)}, \Hc) \mbox{ iff not }
       \abstsatisfied(\vv{x==y}, \Hc)
\]
\end{squeeze}
\caption{Symbolic Execution of Basic Statements}
\label{fig:symExec}
\end{figure*}

\subsubsection{Symbolic Execution} \label{sec:symExec}

Figure~\ref{fig:symExec} shows the symbolic execution
relation $\symexec{st}$.  In most cases, the symbolic
execution of a statement acts on the abstract heap in the
same way that the statement would act on the concrete heap.
In particular, the Store statement always performs strong
updates.  The simplicity of symbolic execution is due to
conditions 3) and 5) in the abstraction relation $\alpha$.
These conditions are ensured by the $\expandeq$ relation
which instantiates nodes, allowing strong updates.  The
symbolic execution also verifies the consistency conditions
that are not verified by $\expandeq$ or $\contracteq$.

\paragraph{Verifying Reference Removal Consistency}
The abstract execution $\transrel{st}$ for the Store
statement can easily verify the Store safety condition
from section \ref{sec:refRemovalCond},
because the set of onstage and offstage nodes is known
precisely for every role graph.  It returns $\Gb$ if the
safety condition fails.

\paragraph{Symbolic Execution of setRole} \label{sec:setRole}
The {\tt setRole(x:r)} statement sets the role of node $n_x$
referenced by variable $\vv{x}$ to $\vv{r}$.  Let $G =
\tu{H,\rho,K}$ be the current role graph and let
$\tu{\abstproc,\vv{x},n_x} \in H$.  If $n_x$ has no adjacent
offstage nodes, the role change always succeeds.  In
general, there are restrictions on when the change can be done.
Let $\tu{\Hc,\rhoc}$ be a concrete heap with role assignment
represented by $G$ and $h$ be a homomorphism from $\Hc$ to
$H$.  Let $h(o_x) = n_x$.  Let $r_0 = \rhoc(o_x)$.  The symbolic execution
must make sure that the condition
$\consistentWith(\rhoc,\Hc,\offstage(\Hc))$ continues to hold
after the role change.  Because the set of onstage nodes
does not change, it suffices to ensure that the original
roles for offstage nodes are consistent with the new role $r$.
The acyclicity constraint involves only offstage nodes, so it
remains satisfied.  The other role constraints are local, so
they can only be violated for offstage neighbors of $n_x$.
To make sure that no violations occur, we require:
\begin{enumerate} \itemsep=0em
\item $\vv{r} \in \field_f(\rho(n))$ for all $\tu{n,f,n_x} \in H$, and
\item $\tu{\vv{r},f} \in \slot_i(\rho(n))$ for all
      $\tu{n_x,f,n} \in H$ and every slot $i$
      such that $\tu{r_0,f} \in \slot_i(\rho(n))$
\end{enumerate}
This is sufficient to guarantee
$\consistentWith(\rhoc,\Hc,\offstage(\Hc))$.  To ensure
condition 2) in Definition~\ref{def:abstrel}
of the abstraction relation, we require that for
every $\tu{f,g} \in \identities(\vv{r})$,
\begin{enumerate}
\item $\tu{f,g} \in \identities(r_0)$ or
\item for all $\tu{n_x,f,n} \in H$: $K(n) = i$ and 
      ($\tu{n,g,n'} \in H$ implies $n'=n_x$).
\end{enumerate}
We use $\rchange(n_x,\vv{r},\tu{H,\rho,K})$ to denote the
check just described.

\paragraph{Symbolic Execution of roleCheck}
To symbolically execute $\vv{roleCheck}(x_1,\ldots,x_p,
\m{ra})$, we ensure that the $\consistentWith$ predicate of
the concrete semantics is satisfied for the concrete heaps
which correspond to the current abstract role graph.  The
symbolic execution for \vv{roleCheck} returns the error
graph $\Gb$ if $\rho$ is inconsistent with $\m{ra}$ or if
any of the nodes $n_i$ referenced by $x_i$ fail to satisfy
$\nodeCheck$.


\subsubsection{Node Check} \label{sec:nodeCheck}

The analysis uses the $\localCheck$, $\acycCheck$,
$\acycCheckAll$, and $\nodeCheck$ predicates to incrementally maintain
the abstraction relation.

We first define the  predicate $\localCheck$, which roughly
corresponds to the predicate $\locallyConsistent$
(Definition
\ref{def:localConsistency}), but ignores the nonlocal acyclicity
condition and additionally ensures condition 2) from
Definition~\ref{def:abstrel}.
\begin{definition} \label{def:localCheck}
For a role graph $G = \tu{H,\rho,K}$, an individual node $n$
and a set $S$, the predicate $\localCheck(n,G)$ holds iff
the following conditions are met.  Let $r = \rho(n)$.
\begin{enumerate} \itemsep=0em
\item[1A.] (Outgoing fields check)
For fields $f \in F$, if $\tu{n,f,n'} \in H$ then
$\rho(n') \in \field_f(r)$.

\item[2A.] (Incoming slots check)
Let $\{\tu{n_1,f_1},\ldots,\tu{n_k,f_k}\} = \{
\tu{n',f} \mid \tu{n',f,n} \in H\}$ be the set of all aliases
of node $n$ in abstract heap $H$. Then $k = \slotno(r)$ and there exists a
permutation $p$ of the set $\{1,\ldots,k\}$ such
that $\tu{\rho(n_i),f_i} \in \slot_{p_i}(r)$ for all $i$.

\item[3A.] (Identity Check)
      If $\tu{n,f,n'} \in H$, $\tu{n',g,n''} \in H$,
      $\tu{f,g} \in \identities(r)$, and $K(n') = i$,
      then $n=n''$.

\item[4A.] (Neighbor Identity Check)
      For every edge $\tu{n',f,n} \in H$, if
      $K(n') = i$, $\rho(n') = r'$ and $\tu{f,g} \in
      \identities(r')$ then $\tu{n,g,n'} \in H$.
\item[5A.] (Field Sanity Check) 
     For every $f \in F$ there is exactly one edge
     $\tu{n,f,n'} \in H$.
\end{enumerate}
\end{definition}
Conditions 1A and 2A correspond to conditions 1) and 2) in
Definition~\ref{def:localConsistency}.  
Condition 3) in Definition~\ref{def:localUptoConsistency} is
not necessarily implied by condition 3A) if
some of the neighbors of $n$ are summary nodes.  Condition
3) cannot be established based only on summary nodes,
because verifying an identity constraint for field $f$ of
node $n$ where $\tu{n,f,n'} \in H$ requires knowing the
identity of $n'$, not only its existence and role.  We
therefore rely on Condition 2) of the
Definition~\ref{def:abstrel} to ensure that identity
relations of neighbors of node $n$ are satisfied before $n$
moves offstage.


The predicate $\acycCheck(n,G,S)$ verifies the acyclicity
condition from Definition~\ref{def:localUptoConsistency} for
the node that has just been brought onstage.  The split
operation uses $\acycCheck$ to check whether it should split
any nodes (Section~\ref{sec:expansion}).  The set $S$
represents offstage nodes.
\begin{definition}
We say that a node $n$ satisfies an acyclicity check in graph $G
= \tu{H,\rho,K}$ with respect to the set $S$, and we write
$\acycCheck(n,G,S)$, iff it is not the case that $H$
contains a cycle $n_1,f_1,\ldots,n_s,f_s,n_1$ where $n_1 =
n$, $f_1,\ldots,f_s \in \acyclic(\rho(n))$ and
$n_1,\ldots,n_s \in S$.
\end{definition}

The analysis uses the predicate $\acycCheckAll(n,G,S)$ in
the contraction relation (Section~\ref{sec:contraction}) to
make sure that $n$ is not a member of any cycle
that would violate the acyclicity condition of any of the nodes
in $S$, including $n$.
\begin{definition}
We say that a node $n$ satisfies a strong acyclicity check in
graph $G = \tu{H,\rho,K}$ with respect to a set $S$, and we
write $\acycCheckAll(n,G,S)$, iff it is not the case that:
$H$ contains a cycle $n_1,f_1,\ldots,n_s,f_s,n_1$ where $n_1
= n$, $f_1,\ldots,f_s \in \acyclic(\rho(n_i))$, for some $1 \leq i
\leq s$, and $n_1,\ldots,n_s \in S$.
\end{definition}
$\acycCheckAll$ is a stronger condition than $\acycCheck$
because it ensures the absence of cycles containing $n$ for
$\acyclic(\rho(n_i))$ fields of all offstage nodes $n_i$,
and not only for the fields $\acyclic(\rho(n))$.

The analysis uses the predicate $\nodeCheck$ to verify that
bringing a node $n$ offstage does not violate role
consistency for offstage nodes.
\begin{definition}
$\nodeCheck(n,G,S)$ holds iff
both predicates $\localCheck(n,G)$ and $\acycCheckAll(n,G,S)$ hold.
\end{definition}



\section{Interprocedural Role Analysis} \label{sec:interAlg}

This section describes the interprocedural aspects of our
role analysis.  Interprocedural role analysis can be viewed
as an instance of the functional approach to interprocedural
data-flow analysis \cite{SharirPnueli81Interprocedural}.
For each program point $p$, role analysis approximates
program traces from procedure entry to point $p$.  The
solution in \cite{SharirPnueli81Interprocedural} proposes
tagging the entire data-flow fact $G$ at point $p$ with the
data flow fact $G_0$ at procedure entry.  In contrast, our
analysis computes the correspondence between heaps at
procedure entry and heaps at point $p$ at the granularity of
sets of objects that constitute role graphs.  This allows
our analysis to detect which regions of the heap have been
modified.  We approximate the concrete executions of a
procedure with {\em procedure transfer relations} consisting
of 1) an initial context and 2) a set of {\em effects}.
Effects are fine-grained transfer relations which summarize
load and store statements and can naturally describe local
heap modifications.  In this paper we assume that procedure
transfer relations are supplied and we are concerned with a)
verifying that transfer relations are a conservative
approximation of procedure implementation b) instantiating
transfer relations at call sites.

\subsection{Procedure Transfer Relations} \label{sec:procTrans}

A transfer relation for a procedure $\m{proc}$ extends the
procedure signature with an initial context
$\context(\m{proc})$, and procedure effects
$\effect(\m{proc})$.

\subsubsection{Initial Context} \label{sec:initialContext}

\begin{figure}[thb]
\begin{center}
\epsfig{file=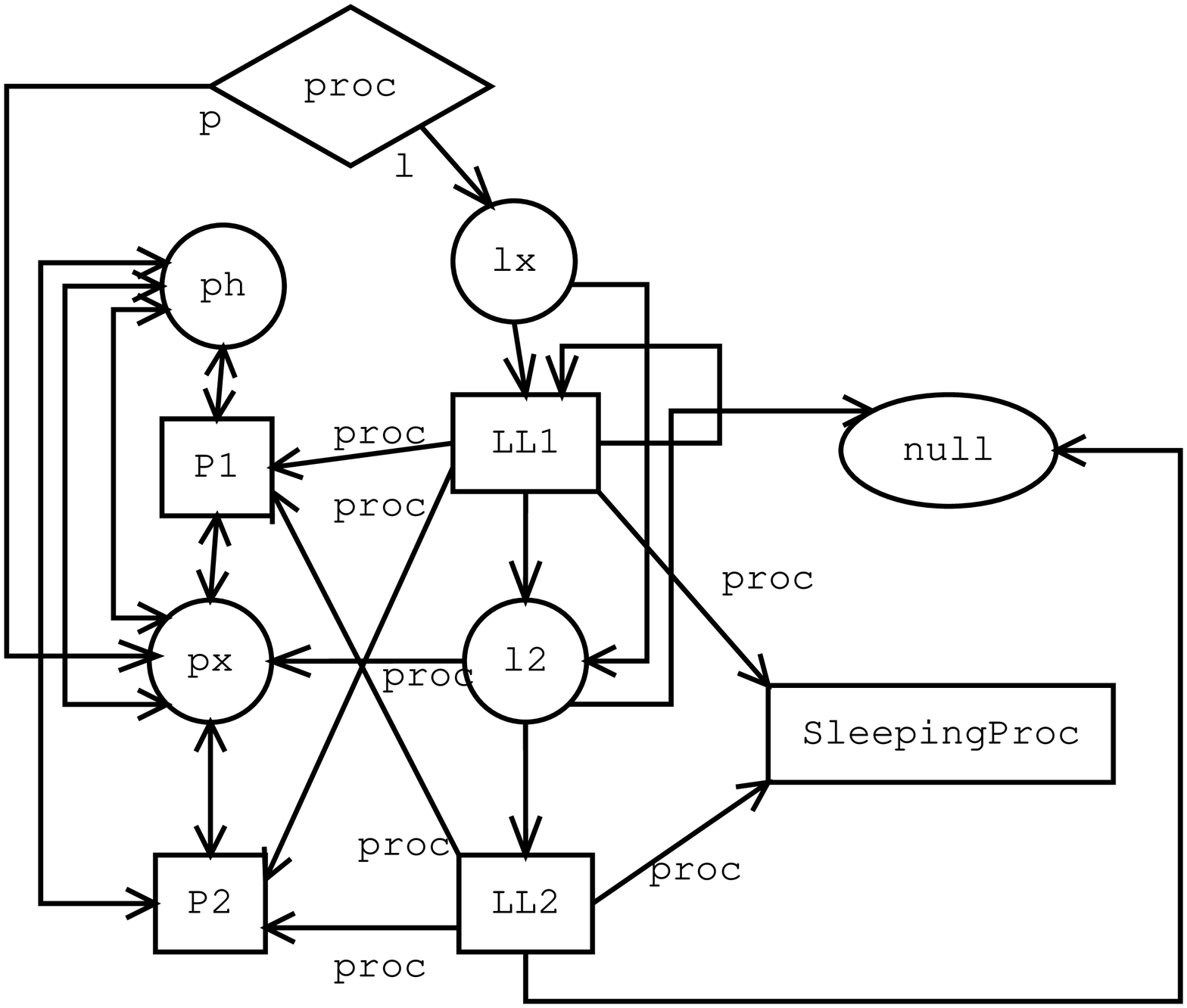,width=3.2in, height=2.5in}
\end{center}
\begin{verbatim}
nodes ph : RunningHeader,
      P1, px, P2 : RunningProc,
      lx : LiveHeader,
      LL1, l2, LL2 : LiveList;
edges p-> px, l-> px,
      ph -next-> P1|px
         -prev-> px|P2,
      P1 -next-> P1|px
         -prev-> ph|P1,
      px -next-> P2|ph
         -prev-> P1|ph,
      P2 -next-> P2|ph
         -prev-> P2|px,
      lx -next-> LL1|l2,
      LL1 -next-> LL1|l2
          -proc-> P1|P2|SleepingProc
      l2 -next-> LL2|null
         -proc-> px,
      LL2 -next-> LL2|null
          -proc-> P1|P2|SleepingProc      
\end{verbatim}
\caption{Initial Context for \vv{kill} Procedure}
\label{fig:killContext}
\end{figure}

Figures \ref{fig:killContext}
and~\ref{fig:insertWithEffects} contain examples of initial
context specification.  An initial context is a description
of the initial role graph $\tu{\Hi,\rhoi,\Ki}$ where $\rhoi$
and $\Ki$ are determined by a $\vv{nodes}$ declaration and
$\Hi$ is determined by a $\vv{edges}$ declaration.  The
initial role graph specifies a set of concrete heaps at
procedure entry and assigns names for sets of nodes in these
heaps.  The next definition is similar to
Definition~\ref{def:abstrel}.
\begin{definition}  \label{def:initContextSemantics}
We say that a concrete heap $\tu{\Hc,\rhoc}$ is represented by
the initial role graph $\tu{\Hi,\rhoi,\Ki}$ and write
$\tu{\Hc,\rhoc} \conabstrel \tu{\Hi,\rhoi,\Ki}$,
iff there exists a function $h_0 :
\nodesc(\Hc) \to \nodes(\Hi)$ such that
\begin{enumerate} \itemsep=0em
\item $\consistentWith(\rhoc,\Hc,h_0^{-1}(\procread(\abstproc))$;
\item $h_0$ is a graph homomorphism;
\item $\Ki(n) = i$ implies $|h_0^{-1}(n)| \leq 1$;
\item $h_0(\nullObj) = \nullAbstNode$ and $h_0(\concproc) = \abstproc$;
\item $\rhoc(o) = \rhoi(h_0(o))$ for every object $o \in
      \nodesc(\Hc)$.
\end{enumerate}
\end{definition}
Here $\procread(\abstproc)$ is the set of initial-context
nodes read by the procedure (see below).  For simplicity, we
assume one context per procedure; it is straightforward to
generalize the treatment to multiple contexts.  

A context is specified by declaring a list of nodes and a
list of edges.

A list of nodes is given with \vv{nodes} declaration.  It
specifies a role for every node at procedure entry.
Individual nodes are denoted with lowercase identifiers,
summary nodes with uppercase identifiers.  By using summary
nodes it is possible to indicate disjointness of entire heap
regions and reachability between nodes in the heap.

There are two kinds
of edges in the initial role graph: parameter edges and heap
edges.  A parameter edge $\vv{p->pn}$ is interpreted as
$\tu{\m{proc},\vv{p},\vv{pn}} \in \Hi$.  We require every
parameter edge to have an individual node as a target, we
call such node a {\em parameter node}.  The role of a
parameter node referenced by $\procparam_i(\m{proc})$ is
always $\prerole_i(\m{proc})$.  
Since different nodes in the initial
role graph denote disjoint sets of concrete objects,
parameter edges
\begin{verbatim}
p1 -> n1
p2 -> n1
\end{verbatim}
imply that parameters $\vv{p1}$ and $\vv{p2}$ must be aliased,
\begin{verbatim}
p1 -> n1
p2 -> n2
\end{verbatim}
force $\vv{p1}$ and $\vv{p2}$ to be unaliased, whereas
\begin{verbatim}
p1 -> n1|n2
p2 -> n1|n2
\end{verbatim}
allow for both possibilities.  A heap edge $\vv{n -f-> m}$
denotes $\tu{\vv{n},\vv{f},\vv{m}} \in \Hi$.  The shorthand
notation
\begin{verbatim}
n1 -f-> n2
   -g-> n3
\end{verbatim}
denotes two heap edges $\tu{\vv{n1},\vv{f},\vv{n2}},
\tu{\vv{n1},\vv{g},\vv{n3}} \in \Hi$.  An expression
$\vv{n1 -f-> n2|n3}$ denotes two edges $\vv{n1 -f-> n2}$ and
$\vv{n1 -f-> n3}$.  We use similar shorthands for parameter
edges.

\begin{example}  \label{exa:killContext}
Figure~\ref{fig:killContext} shows an initial context graph
for the \vv{kill} procedure from
Example~\ref{exa:processOp}.  It is a refinement of the role
reference diagram of Figure~\ref{fig:schedulerSRD} as it
gives description of the heap specific to the entry of
\vv{kill} procedure.  The initial context makes explicit the fact that there 
is only one header node for the list of running
processes (\vv{ph}) and one header node for the list of all
active processes (\vv{lx}).  More importantly, it shows that
traversing the list of active processes reaches a node
\vv{l2} whose \vv{proc} field references the parameter node
\vv{px}.  This is sufficient for the analysis 
to conclude that there will be no null pointer dereferences
in the \vv{while} loop of \vv{kill} procedure since
\vv{l2} is reached before null.
\end{example}

We assume that the initial context always contains the role
reference diagram $\RRD$ (Definition~\ref{def:RRD}).  Nodes
from $\RRD$ are called {\em anonymous nodes} and are
referred to via role name.  This further reduces the size of
initial context specifications by leveraging global role
definitions.  In Figure~\ref{fig:killContext} there is no
need to specify edges originating from \vv{SleepingProc} or
even mention the node \vv{SleepingTree}, since role
definitions alone contain enough information on this part of
the heap to enable the analysis of the procedure.
Note, however, that all edges
between anonymous nodes and named nodes must be explicitly
specified.

\subsubsection{Procedure Effects}

Procedure effects conservatively approximate the region of
the heap that the procedure accesses and indicate changes to
the referencing relationships in that region.  There are two
kinds of effects: read effects and write effects.  

A {\em read effect} specifies a set $\procread(\m{proc})$ of
initial graph nodes accessed by the procedure.  It is used
to ensure that the accessibility condition in
Section~\ref{sec:procCallCons} is satisfied.  If the set of
nodes denoted by $\procread(\m{proc})$ is mapped to a node
$n$ which is onstage in the caller but is not an argument of
the procedure call, a role check error is reported at the
call site.

{\em Write effects} are used to modify caller's role graph
to conservatively model the procedure call.  A write effect
$e_1.f = e_2$ approximates Store operations within a
procedure.  The expression $e_1$ denotes objects being
written to, $f$ denotes the field written, and $e_2$ denotes
the set of objects which could be assigned to the field.
Write effects are {\em may} effects by default, which means
that the procedure is free not to perform them.  It is
possible to specify that a write effect {\em must} be
performed by prefixing it with a ``\vv{!}'' sign.

\begin{figure}
\begin{verbatim}
procedure insert(l : L, 
                 x : IsolatedN ->> LN)
nodes ln, xn;
edges l-> ln, x-> xn,
      ln -next-> LN|null;
effects ln|LN . next = xn,
        ! xn.next = LN|null;
local c, p;
{
  p = l;  
  c = l.next;
  while (c!=null) {
    p = c;  
    c = p.next;
  }
  p.next = x;  
  x.next = c;
  setRole(x:LN);
}
\end{verbatim}
\vspace*{-2em}
\caption{Insert Procedure for Acyclic List}{}
\label{fig:insertWithEffects}
\end{figure}

\begin{example}
In Figure~\ref{fig:insertWithEffects}, the \vv{insert}
procedure inserts an isolated cell into the end of an
acyclic singly linked list.  As a result, the role of the
cell changes to \vv{LN}.  The initial context declares
parameter nodes \vv{ln} and \vv{xn} (whose initial roles are
deduced from roles of parameters), and mentions anonymous
\vv{LN} node from a default copy of the role reference
diagram $\RRD$.  The code of the procedure is summarized
with two write effects.  The first write effect indicates
that the procedure may perform zero or more Store operations
to field
\vv{next} of nodes mapped to \vv{ln} or
\vv{LN} in $\context(\m{proc})$.  The second write
effect indicates that the execution of the procedure must
perform a Store to the field \vv{next} of \vv{xn} node where
the reference stored is either a node mapped onto anonymous
\vv{LN} node or \vv{null}.
\end{example}

Effects also describe assignments that procedures perform on
the newly created nodes.  Here we adopt a simple solution of
using a single summary node denoted $\NEW$ to represent all
nodes created inside the procedure.  We write
$\nodes_0(\Hi)$ for the set $\nodes(\Hi) \cup \{\NEW\}$.

\begin{figure}
\begin{verbatim}
procedure insertSome(l : L)
nodes ln;
edges l-> ln,
      ln -next-> LN|null;
effects ln|LN . next = NEW,
        NEW.next = LN|null;
aux c, p, x;
{
  p = l;  
  c = l.next;
  while (c!=null) {
    p = c;  
    c = p.next;
  }
  x = new;
  p.next = x;  
  x.next = c;
  setRole(x:LN);
}
\end{verbatim}
\caption{Insert Procedure with Object Allocation}
\label{fig:insertSome}
\end{figure}

\begin{example}
Procedure \vv{insertSome} in Figure~\ref{fig:insertSome} is
similar to procedure \vv{insert} in
Figure~\ref{fig:insertWithEffects}, except that the node
inserted is created inside the procedure.  It is therefore
referred to in effects via generic summary node $\NEW$.
\end{example}

We represent all may write effects as a set
$\procmaywrite(\m{proc})$ of triples $\tu{n_j,f,n'_j}$ where
$n,n'_j \in \nodes_0(\Hi)$ and $f \in F$.  We represent must
write effects as a sequence $\procmustwrite_j(\m{proc})$ of
subsets of the set $\Ki^{-1}(i) \times F \times
\nodes_0(\Hi)$.  Here $1
\leq j \leq \mustwriteno(\m{proc})$.

To simplify the interpretation of the declared procedure
effects in terms of concrete reads and writes, we require
the union $\cup_i \procmustwrite_i(\m{proc})$ to be disjoint
from the set $\procmaywrite(\m{proc})$.  We also require the
nodes $n_1,\ldots,n_k$ in a must write effect
$n_1|\cdots|n_k . f = e_2$ to be individual nodes.  This
allows strong updates when instantiating effects
(Section~\ref{sec:effectInstantiation}).

\subsubsection{Semantics of Procedure Effects}

We now give precise meaning to procedure effects.  Our
definition is slightly complicated by the desire to capture
the set of nodes that are actually read in an execution
while still allowing a certain amount of observational
equivalence for write effects.

The effects of procedure $\m{proc}$ define a subset of
permissible program traces in the following way.  Consider a
concrete heap $\Hc$ with role assignment $\rhoc$ such that
$\tu{\Hc,\rhoc} \conabstrel
\tu{\Hi,\rhoi,\Ki}$ with graph homomorphism $h_0$ from
Definition~\ref{def:initContextSemantics}.  Consider a trace
$T$ starting from a state with heap $\Hc$ and role
assignment $\rhoc$.  Extract the subsequence of all loads
and stores in trace $T$.  Replace Load \vv{x=y.f} by
concrete read $\vv{read } o_x$ where $o_x$ is the concrete
object referenced by $\vv{x}$ at the point of Load, and
replace Store \vv{x.f=y} by a concrete write $o_x.f = o_y$
where $o_x$ is the object referenced by
\vv{x} and $o_y$ object referenced by \vv{y} at the point of
Store.  Let $p_1,\ldots,p_k$ be the sequence of all concrete
read statements and $q_1,\ldots,q_k$ the sequence of all
concrete write statements.  We say that trace $T$ starting
at $\Hc$ conforms to the effects iff for all choices of
$h_0$ the following conditions hold:
\begin{enumerate} \itemsep=0em
\item $h_0(o) \in \procread(\m{proc})$ for every $p_i$ of the form 
      $\vv{read } o$
\item there exists a subsequence $q_{i_1},\ldots,q_{i_t}$ 
      of $q_1,\ldots,q_k$ such that
\begin{enumerate}
\item executing $q_{i_1},\ldots,q_{i_t}$ on $\Hc$ yields the
      same result as executing the entire sequence $q_1,\ldots,q_k$
\item the sequence $q_{i_1},\ldots,q_{i_t}$
      implements write effects of procedure $\m{proc}$
\end{enumerate}
\end{enumerate}
A typical way to obtain a sequence $q_{i_1},\ldots,q_{i_t}$
from the sequence $q_1,\ldots,q_k$ is to consider only the
last write for each pair $\tu{o_i,f}$ of object and field.

We say that a sequence $q_{i_1},\ldots,q_{i_t}$ implements
write effects $\procmaywrite(\m{proc})$ and
$\procmustwrite_i(\m{proc})$ for $1 \leq i \leq i_0$, $i_0 =
\mustwriteno$ if and only if there exists an injection $s :
\{1,\ldots,i_0\} \to \{i_1,\ldots,i_t\}$ such that
\begin{enumerate}
\item $\tu{h'(o),f,h'(o')} \in \procmustwrite_i(\m{proc})$ for every
      concrete write $q_{s(i)}$ of the form $o.f=o'$, and
\item $\tu{h'(o),f,h'(o')} \in \procmaywrite(\m{proc})$ for
      all concrete writes $q_i$ of the form $o.f = o'$ for
      $i \in \{i_1,\ldots,i_t\} \setminus \{s(1),\ldots,s(i_0)\}$.
\end{enumerate}
Here $h'(n) = h_0(n)$ for $n \in \nodes(\Hc)$ where $\Hc$ is
the initial concrete heap and $h'(n) = \NEW$ otherwise.

It is possible (although not very common) for a single
concrete heap $\Hc$ to have multiple homomorphisms $h_0$ to
the initial context $\Hi$.  Note that in this case we
require the trace $T$ to conform to effects for {\em all}
possible valid choices of $h_0$.  This places the burden of
multiple choices of $h_0$ on procedure transfer relation
verification (Section~\ref{sec:verifyingTransRel}) but in
turn allows the context matching algorithm in
Section~\ref{sec:contextMatching} to select an arbitrary
homomorphism between a caller's role graph and an initial
context.

\subsection{Verifying Procedure Transfer Relations}
\label{sec:verifyingTransRel}

In this section we show how the analysis makes sure that a
procedure conforms to its specification, expressed as an
initial context with a list of effects.  To verify procedure
effects, we extend the analysis representation from
Section~\ref{sec:anaAbstraction}.  A non-error role graph is
now a tuple $\tu{H,\rho,K,\tau,E}$ where:
\begin{enumerate} \itemsep=0em
\item $\tau : \nodes(H) \to \nodes_0(\Hi)$
      is initial context transformation that assigns an
      initial context node $\tau(n) \in \nodes(\Hi)$ to
      every node $n$ representing objects that
      existed prior to the procedure call, and assigns
      $\NEW$ to every node representing objects
      created during procedure activation;
\item $E \subseteq \cup_i \procmustwrite_i(\m{proc})$ is a list
      of must write effects that procedure has performed so far.
\end{enumerate}
The initial context transformation $\tau$ tracks how objects have
moved since the beginning of procedure activation and is
essential for verifying procedure effects which refer to
initial context nodes.

We represent the list $E$ of performed must effects as a
partial map from the set $\Ki^{-1}(i) \times F$ to
$\nodes_0(\Hi)$.  This allows the analysis to perform
must effect folding by recording only the last must effect
for every pair $\tu{n,f}$ of individual node $n$ and field
$f$.

\begin{figure}[thb]
\[\begin{array}{r@{}l}
\tr{\vv{entry}\bullet} = \Big\{
& \tu{H,\rho,K,\tau,E} \>\Big|\> \\
& P : \{\abstproc\} \times \{\procparam_i(\abstproc)\}_i \to N, P \subseteq \Hi \\
& H_0 = (\Hi \setminus \{\abstproc\} \times \procparam(\abstproc) \times N) \cup P \\
& n_i = P(\abstproc,\procparam_i(\abstproc)) \\
& H_1 \subseteq H_0 \\
& H_1 \setminus H_0 \subseteq \{ \tu{n',f,n''} \mid 
       \{n_1,n_2\} \cap \{n_i\}_i \neq \emptyset \} \\
& \forall j : \localCheck(n_j,\tu{H,\rho,K},\nodes(H_1)) \\
& H_1 \splt{n_1} H_2 \splt{n_2} {} \cdots {} \splt{n_p} H \\
& \rho = \rhoi \\
& K = \Ki \\
& \tau = \rhoi \\
& E = \emptyset \> \Big\}\\
\end{array}\]
\caption{The Set of Role Graphs at Procedure Entry}
\label{fig:creatingInit}
\end{figure}

\subsubsection{Role Graphs at Procedure Entry} \label{sec:roleGraphsOnEntry}

Our role analysis creates the set of role graphs at procedure
entry point from the initial context $\context(\m{proc})$.
This is simple because role graphs and the initial
context have similar abstraction relations (Sections
\ref{sec:anaAbstraction} and~\ref{sec:procTrans}).
The difference is that parameters in role graphs point to
exactly one node, and parameter nodes are onstage nodes in
role graphs which means that all their edges are ``must''
edges.

Figure~\ref{fig:creatingInit} shows the construction of the
initial set of role graphs.  First the graph $H_0$ is created
such that every parameter $\procparam_i(\m{proc})$ references
exactly one parameter node $n_i$.  Next graph $H_1$ is
created by using $\localCheck$ to ensure that parameter
nodes have the appropriate number of edges.  Finally, the
instantiation is performed on parameter nodes to ensure
acyclicity constraints if the initial context does not make
them explicit already.

\begin{figure*}
\newcommand{\transi}[2]{{#1} \symexec{s} {#2}}
\begin{squeeze}
\[\begin{array}{|c|c|c|} \hline
\mbox{Statement } $s$
&
\mbox{Transition}
&
\mbox{Constraints}
\\ \hline \hline
\vv{x = y.f}
&
\transi{\tu{H \withp{\abstproc,\vv{x},n_x},\rho,K,\tau,E}}
       {\tu{H \withp{\abstproc,\vv{x},n_f},\rho,K,\tau,E}}
&
\vstack{
   \tu{\abstproc,\vv{y},n_y}, \tu{n_y,f,n_f} \in H \\
   \tau(n_f) \in \procread(\abstproc)
}
\\ \hline
\vv{x = y.f}
&
\transi{\tu{H \withp{\abstproc,\vv{x},n_x},\rho,K,\tau,E}}
       {\Gb}
&
\vstack{
   \tu{\abstproc,\vv{y},n_y}, \tu{n_y,f,n_f} \in H \\
   \tau(n_f) \notin \procread(\abstproc)
}
\\ \hline
\vv{x.f = y}
&
\transi{\tu{H \withp{n_x,f,n_f},\rho,K,\tau,E}}
       {\tu{H \withp{n_x,f,n_y},\rho,K,\tau,E}}
&
\vstack {
   \tu{\abstproc,\vv{x},n_x},\tu{\abstproc,\vv{y},n_y} \in H \\
   \tu{\tau(n_x),f,\tau(n_y)} \in \procmaywrite(\abstproc) \\
}
\\ \hline
\vv{x.f = y}
&
\transi{\tu{H \withp{n_x,f,n_f},\rho,K,\tau,E}}
       {\tu{H \withp{n_x,f,n_y},\rho,K,\tau,E'}}
&
\vstack {
   \tu{\abstproc,\vv{x},n_x},\tu{\abstproc,\vv{y},n_y} \in H \\
   \tu{\tau(n_x),f,\tau(n_y)} \in \cup_i \procmustwrite_i(\abstproc) \\
   E' = \updateMustWrite(E, \tu{\tau(n_x),f,\tau(n_y)}) \\
}
\\ \hline
\vv{x.f = y}
&
\transi{\tu{H \withp{n_x,f,n_f},\rho,K,\tau,E}}
       {\Gb}
&
\vstack {
   \tu{\abstproc,\vv{x},n_x},\tu{\abstproc,\vv{y},n_y} \in H \\
   \vstack { \tu{\tau(n_x),f,\tau(n_y)} \notin 
              \procmaywrite(\abstproc) \cup \\
              \cup_i \procmustwrite_i(\abstproc)
           } \\
}
\\ \hline
\vv{x = new}
&
\transi{\tu{H \withp{\abstproc,\vv{x},n_x},\rho,K,\tau,E}}
           {\tu{H \withp{\abstproc,\vv{x},n_n},\rho,K,\tau',E}}
&
\vstack{
         n_n \mbox{ fresh} \\
         \tau' = \tau[n_n \mapsto \NEW] \\
       }
\\ \hline
\end{array}\]
\[
    \updateMustWrite(E, \tu{n_1,f,n_2}) = E[\tu{n_1,f} \mapsto n_2]
\]
\end{squeeze}
\caption{Verifying Load, Store, and New Statements}{}
\label{fig:effectVerif}
\end{figure*}

\subsubsection{Verifying Basic Statements}

To ensure that a procedure conforms to its transfer relation
the analysis uses the initial context transformation $\tau$
to assign every Load and Store statement to a declared
effect.  Figure~\ref{fig:effectVerif} shows new symbolic
execution of Load, Store and New statements.  

The symbolic execution of Load statement \vv{x=y.f} makes
sure that the node being loaded is recorded in some read
effect.  If this is not the case, an error is reported.  

The symbolic execution of the Store statement
\vv{x.f=y} first retrieves nodes $\tau(n_x)$ and
$\tau(n_y)$ in the initial role graph $\context$ that
correspond to nodes $n_x$ and $n_y$ in the current role
graph.  If the effect $\tu{\tau(n_x),f,\tau(n_y)}$ is
declared as a may write effect the execution proceeds as
usual.  Otherwise, the effect is used to update the list $E$
of must-write effects.  The list $E$ is checked at the end
of procedure execution.  

The symbolic execution of the New statement updates the
initial context transformation $\tau$ assigning $\tau(n_n) =
\NEW$ for the new node $n_n$.  

The $\tau$ transformation is similarly updated during other
abstract heap operations.  Instantiation of node $n'$ into
node $n_0$ assigns $\tau(n_0) = \tau(n')$, split copies
values of $\tau$ into the new set of isomorphic nodes, and
normalization does not merge nodes $n_1$ and $n_2$ if
$\tau(n_1) \neq \tau(n_2)$.

\subsubsection{Verifying Procedure Postconditions}

At the end of the procedure, the analysis verifies that
$\rho(n_i) = \postrole_i(\m{proc})$ where 
$\tu{\abstproc,\procparam_i(\m{proc}),n_i} \in H$, and then
performs node check on all onstage nodes using predicate
$\nodeCheck(n,\tu{H,\rho,K},\nodes(H))$ for all $n \in
\onstage(H)$.  

At the end of the procedure, the analysis also verifies that
every performed effect in $E = \{e_1,\ldots,e_k\}$ can be
attributed to exactly one declared must effect.  This means
that $k = \mustwriteno(\m{proc})$ and there exists a
permutation $s$ of set $\{1,\ldots,k\}$ such that $e_{s(i)}
\in \procmustwrite_i(\m{proc})$ for all $i$.

\begin{figure}[bth]
\[\begin{array}{l}
   \tr{\abstproc'(x_1,\ldots,x_p)}(\GS) = \\
\quad
\begin{array}{l}
\xif \exists G \in \GS : \lnot \paramCheck(G) \xthen \{\Gb\} \\
  \begin{array}{l@{}l}
\xelse & \m{try } \GS_1 = \matchContext(\GS) \\
  & \xif \m{failed} \xthen \{\Gb\} \\
  & \begin{array}{l@{}l}
    \xelse \{ & G'' \mid \tu{G,\mu} \in \GS_1 \\
                     & \tu{\addNEW(G),\mu} \doeffects \tu{G',\mu} \doroles G'' \} \\
           \end{array} \\
  \end{array} \\
\end{array} \\ 
\\
\begin{array}{l}
\paramCheck(\tu{H,\rho,K,\tau,E}) \mbox{ iff } \\
\qquad \forall n_i : \nodeCheck(n_i,G,\offstage(H) \cup \{n_i\}_i) \\
\qquad n_i \mbox{ are such that } \tu{\abstproc,x_i,n_i} \in H \\
\end{array} \\
\\
\begin{array}{l}
\addNEW(\tu{H,\rho,K,\tau,E}) = \\
\qquad      \tu{H \cup \{n_0\} \times F \times \{\nullAbstNode\}, \\
\qquad      \rho[n_0 \mapsto \unknownRole], \\
\qquad      K[n_0 \mapsto s], \\
\qquad      \tau[n_0 \mapsto \NEW], \\
\qquad      E} \\
\mbox{where } n_0 \mbox{ is fresh in } H \\
\end{array}
\end{array}\]
\caption{Procedure Call}
\label{fig:procCallTransfer}
\end{figure}

\subsection{Analyzing Call Sites}

The set of role graphs at the procedure call site is updated
based on the procedure transfer relation as follows.  Consider
procedure $\abstproc$ containing call site $p \in
\CFGN(\abstproc)$ with procedure call
$\abstproc'(x_1,\ldots,x_p)$.  Let $\tu{\Hi,\rhoi,\Ki} =
\context(\abstproc')$ be the initial context of the callee.

Figure~\ref{fig:procCallTransfer} shows the transfer
function for procedure call sites.  It has the following
phases:
\begin{enumerate} \itemsep=0em
\item {\bf Parameter Check} ensures that roles of
      parameters conform to the roles expected by the callee $\abstproc'$.
\item {\bf Context Matching} ($\matchContext$) ensures that the 
      caller's role graphs
      represent a subset of concrete heaps represented by
      $\context(\abstproc')$.  This is done by deriving a
      mapping $\mu$ from the caller's role graph to
      $\nodes(\Hi)$.
\item {\bf Effect Instantiation} ($\doeffects{}{}{}$) uses effects
      $\procmaywrite(\abstproc')$ and $\procmustwrite_i(\abstproc')$
      in order to approximate all structural changes to the role graph that
      $\abstproc'$ may perform.
\item {\bf Role Reconstruction} ($\doroles$) uses final 
      roles for parameter nodes and
      global role declarations $\postrole_i(\abstproc')$
      to reconstruct roles of all
      nodes in the part of the role graph representing
      modified region of the heap.
\end{enumerate}
The parameter check requires $\nodeCheck(n_i,G,\offstage(H) \cup
\{n_i\}_i)$ for the parameter nodes $n_i$.
The other three phases are explained in more detail below.

\subsubsection{Context Matching} \label{sec:contextMatching}

\begin{figure*}[thb]
\[
    \matchContext(\GS) = 
         \match(\{\tu{G,\nodes(G)\times \{\bot\}} \mid G \in \GS \})
\]
\[\begin{array}{l}
\match : {\cal P}(\RoleGraphs \times (N\cup\{\bot\})^N) \rightharpoonup
         {\cal P}(\RoleGraphs \times N^N) \\
\\
\match(\GM) = {}\\
\begin{array}{l@{}ll}
& \GM_0 := \{ \tu{G,\mu} \in \GM \mid \mu^{-1}(\bot) \neq \emptyset \}; \\
& \xif \GM_0 = \emptyset \xthen \return \GM; \\
& \tu{\tu{H,\rho,K,\tau,E},\mu} := \xchoose \GM_0; \\
& \GM' = \GM \setminus \tu{\tu{H,\rho,K,\tau,E},\mu}; \\
& \theseparams := \{ n \mid \exists i : \tu{\abstproc,x_i,n} \in H \}; \\
& \inaccessible := \onstage(H) \setminus \theseparams; \\
& n_0 := \xchoose \mu^{-1}(\bot); \\
& \candidates := \{ n' \in \nodes(\Hi) \mid \\
& \begin{array}{r@{}l}
                & \qquad\qquad (n_0 \notin \inaccessible \mbox{ and } \rhoi(n') = \rho(n_0)) \mbox{ or } \\
                & \qquad\qquad (n_0 \in \inaccessible \mbox{ and } n' \notin \procread(\abstproc')) \} \\
                & {} \bigcap\limits_{\vstack{
                                         \scriptstyle \tu{n_0,f,n} \in H \\
                                         \scriptstyle \mu(n) \neq \bot
                                     }}
                        \Big\{ n' \>\Big|\> \tu{n',f,\mu(n)} \in \Hi \Big\} \\
                & {} \bigcap\limits_{\vstack{
                                         \scriptstyle \tu{n,f,n_0} \in H \\
                                         \scriptstyle \mu(n) \neq \bot
                                     }}
                        \Big\{ n'  \>\Big|\> \tu{\mu(n),f,n'} \in \Hi \Big\}; \\
  \end{array} \\
& \xif \candidates = \emptyset \xthen \xfail; \\
& \xif \candidates = \{ n'_0 \}, K(n_0) = s, \Ki(n'_0) = i, \mu^{-1}(n'_0) = \emptyset \\
&  \quad \xthen \match(\GM' \cup \{ \tu{G', \mu[n_1 \mapsto n'_0]} \mid 
                       \tu{H,\rho,K,\tau,E} \instantiate{n_0}{n_1} G' \}) \\
& \begin{array}{l@{}l}
    \xelse &   n'_0 := \xchoose \{ n' \in \candidates \mid K(n') = s \mbox{ or } \\
           & \qquad\qquad\qquad  (K(n_0) = i, \mu^{-1}(n') = \emptyset) \} \\
           & \match(\GM' \cup \tu{\tu{H,\rho,K,\tau,E},\mu[n_0 \mapsto n'_0]}); \\
  \end{array}
\end{array}
\end{array}\]
\caption{The Context Matching Algorithm}
\label{fig:matching}
\end{figure*}

Figure~\ref{fig:matching} shows our context matching
function. The $\matchContext$ function takes a set
$\GS$ of role graphs and produces a set of pairs
$\tu{G,\mu}$ where $G = \tu{H,\rho,K,\tau,E}$ is a role
graph and $\mu$ is a homomorphism from $H$ to $\Hi$.  The
homomorphism $\mu$ guarantees that $\alpha^{-1}(G) \subseteq
\alpha_0^{-1}(\context(\abstproc'))$ since the homomorphism
$h_0$ from Definition~\ref{def:initContextSemantics} can be
constructed from homomorphism $h$ in
Definition~\ref{def:abstrel} by putting $h_0 = \mu \circ h$.
This implies that it is legal to call $\abstproc'$ with any
concrete graph represented by $G$.

The algorithm in Figure~\ref{fig:matching} starts with empty
maps $\mu = \nodes(G) \times \{\bot\}$ and extends $\mu$
until it is defined on all $\nodes(G)$ or there is no way
to extend it further.  It proceeds by choosing a role graph
$\tu{H,\rho,K,\tau,E}$ and node $n_0$ for which the mapping $\mu$
is not defined yet.  It then finds candidates in the initial
context that $n_0$ can be mapped to.  The candidates are
chosen to make sure that $\mu$ remains a homomorphism.
The accessibility requirement---that a procedure may see no
nodes with incorrect role---is enforced by making sure that
nodes in $\inaccessible$ are never mapped into nodes in
$\procread$ for the callee.  As long as this requirement
holds, nodes in $\inaccessible$ can be mapped onto nodes of
any role since their role need not be correct anyway.  We
generally require that the set $\mu^{-1}(n'_0)$ for
individual node $n'_0$ in the initial context contain at
most one node, and this node must be individual.  In
contrast, there might be many individual and summary nodes
mapped onto a summary node.  We relax this requirement
by performing instantiation of a summary node of the caller
if, at some point, that is the only way to extend the
mapping $\mu$ (this corresponds to the first recursive call
in the definition of $\match$ in Figure~\ref{fig:matching}).

The algorithm is nondeterministic in the order in which
nodes to be matched are selected.  One possible ordering of
nodes is depth-first order in the role graph starting from
parameter nodes.  If some nondeterministic branch does not
succeed, the algorithm backtracks.  The function fails if
all branches fail.  In that case the procedure call is
considered illegal and $\Gb$ is returned.  The algorithm
terminates since every procedure
call lexicographically increases the sorted list of numbers
$|\mu[\nodes(H)]|$ for $\tu{\tu{H,\rho,K,\tau,E},\mu}
\in \GM$.

%
%

\subsubsection{Effect Instantiation} \label{sec:effectInstantiation}

The result of the matching algorithm is a set of pairs
$\tu{G,\mu}$ of role graphs and mappings.  These pairs are
used to instantiate procedure effects in each of the role
graphs of the caller.  Figure~\ref{fig:effectInst} gives
rules for effect instantiation.  The analysis first verifies
that the region read by the callee is included in the region
read by the caller.  Then it uses map $\mu$ to find the
inverse image $S$ of the performed effects.  The effects in
$S$ are grouped by the source $n$ and field $f$.  Each field
$n.f$ is applied in sequence.  There are three cases when
applying an effect to $n.f$:
\begin{enumerate} \itemsep=0em
\item There is only one node target of the write
      in $\nodes(H)$ and the effect is a must write effect.
      In this case we do a strong update.
\item The condition in 1) is not satisfied, and the node $n$
      is offstage.  In this case we conservatively add all
      relevant edges from $S$ to $H$.
\item The condition in 1) is not satisfied, but the node $n$
      is onstage i.e. it is a parameter
      node\footnote{Non-parameter onstage nodes are never
      affected by effects, as guaranteed by the matching
      algorithm.}.  
       In this case there is no unique target
       for $n.f$, and we cannot add multiple edges either as
       this would violate the invariant for onstage nodes.
      We therefore do case analysis choosing which effect
      was performed last.  If there are no must effects that
      affect $n$, then we also consider the case where the
      original graph is unchanged.
\end{enumerate}

\subsubsection{Role Reconstruction}  \label{sec:roleReconstruction}

\begin{figure*}[tbh]
\[
   \tu{\tu{H,\rho,K,\tau,E},\mu} \doroles \tu{H',\rho',K',\tau',E'}
\]
\[\begin{array}{l}
\tu{\abstproc,x_i,n_i} \in H \\
N_0 = \mu^{-1}[\procread(\abstproc')] \\
s : N_0 \times R \to N \mbox{ where $s(n,r)$ are all different nodes fresh in $H$ } \\
\rho' = \rho \setminus (N_0 \times R) \cup \{ \tu{s(n,r),r} \mid n \in N_0, r \in R \} \\
 \qquad \qquad \setminus (\{n_i\}_i \times R) \cup \{ \tu{n_i,\postrole_i(\abstproc)} \} \\
K'(s(n,r)) = K(n) \\
\tau'(s(n,r)) = \tau(n) \\
E' = E \\
H_0 = H \setminus \{ \tu{n_1,f,n_2} \mid n_1 \in N_0 \mbox{ or } n_2 \in N_0 \} \\
\qquad \qquad {} \cup \{ \tu{s(n_1,r_1),f,s(n_2,r_2)} \mid \tu{n_1,f,n_2} \in H, 
                                                   \tu{r_1,f,r_2} \in \RRD \} \\
\qquad \qquad {} \cup \{ \tu{n_1,f,s(n_2,r_2)} \mid \tu{n_1,f,n_2} \in H, 
                                       \tu{\rhoi(\mu(n_1)),f,r_2} \in \RRD \} \\
\qquad \qquad {} \cup \{ \tu{s(n_1,r_1),f,n_2} \mid \tu{n_1,f,n_2} \in H,
                                       \tu{r_1,f,\rhoi(\mu(n_2))} \in \RRD \} \\
H' = \GC(H_0) \\
\end{array}\]
\caption{Call Site Role Reconstruction}
\label{fig:roleRecons}
\end{figure*}

Procedure effects approximate structural changes to the
heap, but do not provide information about role changes for
non-parameter nodes.  We use the role reconstruction
algorithm $\doroles$ in Figure~\ref{fig:roleRecons} to
conservatively infer possible roles of nodes after the
procedure call based on role changes for parameters and
global role definitions.

Role reconstruction first finds the set $N_0$ of all
nodes that might be accessed by the callee since these nodes
might have their roles changed.  Then it splits each node $n
\in N_0$ into $|R|$ different nodes $\rho(n,r)$, one for
each role $r \in R$.  The node $\rho(n,r)$ represents the
subset of objects that were initially represented by $n$ and
have role $r$ after procedure executes.
The edges between nodes in the new graph are derived
by simultaneously satisfying 1) structural constraints
between nodes of the original graph; and 2) global role
constraints from the role reference diagram.  
The nodes $\rho(n,r)$ not connected to the parameter nodes
are garbage collected in the role graph.  In practice, we
generate nodes $\rho(n,r)$ and edges on demand
starting from parameters making sure that they are
reachable and satisfy both kinds of constraints.

\section{Extensions} \label{sec:extensions}

This section presents two extensions of the basic role
system.  The first extension allows statically unbounded
number of aliases for objects.  The second extension allows
the analysis to verify more complex role changes.
Additional ways of extending roles are given in
\cite{Kuncak01DesigningRoleAnalysis}.

\subsection{Multislots}

A multislot $\tu{r',f} \in \multislots(r)$ in the definition
of role $r$ allows any number of aliases $\tu{o',f,o} \in
\Hc$ for $\rhoc(o') = r'$ and $\rhoc(o) = r$.  We
require multislots $\multislots(r)$ to be disjoint from all
$\slot_i(r)$.  To handle multislots in role analysis we
relax the condition 5) in Definition~\ref{def:abstrel} of
the abstraction relation by allowing $h$ to map more than
one concrete edge $\tu{o',f,o}$ onto abstract edge
$\tu{n',f,n}
\in H$ terminating at an onstage node $n$ provided that
$\tu{\rho(n'),f} \in \multislots(\rho(n))$.  The
$\nodeCheck$ and expansion relation $\expandeq{}$ are then
extended appropriately.  Note that a role graph does not
represent the exact number of references that fill each
multislot. The analysis therefore does not attempt to
recognize actions that remove the last reference from the
multislot. Once an object plays a role with a multislot, all
subsequent roles that it plays must also have the multislot.

\subsection{Cascading Role Changes} \label{sec:cascadingChange}

In some cases it is desirable to change roles of an entire
set of offstage objects without bringing them onstage.  We
use the statement
$\vv{setRoleCascade}(x_1:r_1,\ldots,x_n:r_n)$ to perform
such {\em cascading role change} of a set of nodes.  The
need for cascading role changes arises when roles encode
reachability properties.

\begin{figure}
\begin{verbatim}
role BufferNode {
  fields next : BufferNode | null;
  slots  BufferNode.next | Data.buffer;
  acyclic next;
}
role WorkNode {
  fields next : WorkNode | null;
         WorkNode.next | Data.work;
  acyclic next;
}
role Data {
  fields buffer : BufferNode | null,
         work   : WorkNode | null;
}

{
   ...
   roleCheck(m : Data);
   x = m.buffer;
   y = m.work;
   m.buffer = y;
   m.work = x;
   setRoleCascade(x:WorkNode, y:BufferNode);
   ...
}
\end{verbatim}
\caption{Example of a Cascading Role Change}
\label{fig:cascadingRoleChange}
\end{figure}

\begin{example}
The code fragment in Figure~\ref{fig:cascadingRoleChange}
manipulates object \vv{m} of role \vv{Data}.
The role \vv{Data} has fields \vv{buffer} and \vv{work}, each
being a root for a singly linked acyclic list.  Elements
of the first list have \vv{BufferNode} role and elements
of the second list have \vv{WorkNode} role.  At some
point procedure swaps the contents of the fields \vv{buffer} and
\vv{work}, which requires all nodes in both lists to change
the roles.  These role changes are triggered by the
\vv{setRoleCascade} statement.  The statement indicates new
roles for onstage nodes, and the analysis cascades role
changes to offstage nodes.
\end{example}

\begin{figure*}[bth]
\begin{squeeze}
\[\begin{array}{|c|c|} \hline
\begin{array}{l}
  \tu{H,\rho,K,\tau,E} \transrel{s} \tu{H,\rho',K,\tau,E} \\
  s = \vv{setRoleCascade}(x_1:r_1,\ldots,x_n:r_n) \\
\end{array}
&
\begin{array}{l}
n_i : \tu{\abstproc,x_i,n_i} \in H \\
\rho'(n_i) = r_i \\
\rho'(n) = \rho(n), \> n \in \onstage(H) \setminus \{n_i\}_i \\
N_0 = \{ n \in \offstage(H) \mid \exists n' \in \neighbors(n,H) :
         \rho(n') \neq \rho'(n') \} \\
\forall n \in N_0 : \cascadingOk(n,H,\rho,K,\rho') \\
\end{array} \\ \hline
\end{array}\]
\end{squeeze}
\caption{Abstract Execution for \vv{setRoleCascade}}
\label{fig:setRoleCascade}
\end{figure*}

Given a role graph $\tu{H,\rho,K,E}$ cascading role change
finds a new valid role assignment $\rho'$ where the onstage
nodes have desired roles and the roles of offstage nodes are
adjusted appropriately.  Figure~\ref{fig:setRoleCascade}
shows abstract execution of the \vv{setRoleCascade}
statement.  Here $\neighbors(n,H)$ denotes nodes in $H$
adjacent to $n$.  The condition
$\cascadingOk(n,H,\rho,K,\rho')$ makes sure it is legal to
change the role of node $n$ from $\rho(n)$ to $\rho'(n)$
given that the neighbors of $n$ also change role according
to $\rho'$.  This check resembles the check for \vv{setRole}
statement in Section~\ref{sec:setRole}.  Let $r = rho(n)$
and $r' = \rho'(n)$.  Then $\cascadingOk(n,H,\rho,K,\rho')$
requires the following conditions:
\begin{enumerate} \itemsep=0em
\item $\tu{n,f,n_1} \in H$ implies $\rho'(n_1) \in \field_f(r')$
\item $\slotno(r') = \slotno(r) = k$, and
      for every list $\tu{n_1,f_1,n}, \ldots,\tu{n_k,f_k,n}
      \in H$ if there is a permutation $p : \{1,\ldots,k\}
      \to \{1,\ldots,k\}$ such that $\tu{\rho(n_i),f_i} \in
      \slot_{p_i}(r)$, then there is a permutation $p' :
      \{1,\ldots,k\} \to \{1,\ldots,k\}$ such that
      $\tu{\rho(n_i),f_i} \in \slot_{p_i}(r')$.
\item identity relations were already satisfied
      or can be explicitly checked: $\tu{f,g} \in
      \identities(\rho'(n))$ implies
 \begin{enumerate} \itemsep=0em
 \item $\tu{f,g} \in \identities(\rho(n))$ or 
 \item for all $\tu{n,f,n'} \in H$:
       $K(n') = i$, and \\ if $\tu{n',g,n''} \in H$ then
       $n''=n$
 \end{enumerate}
\item either $\acyclic(\rho'(n)) \subseteq \acyclic(\rho(n))$ 
      or \\ $\acycCheck(n,\tu{H,\rho',K},\offstage(H))$.
\end{enumerate}
In practice there may be zero or more solutions that satisfy
constraints for a given cascading role change.  Selecting
any solution that satisfies the constraints is sound with
respect to the original semantics.  A useful heuristic for
searching the solution space is to first explore branches
with as few roles changed as possible.  If no solutions are
found, an error is reported.

\section{Related Work}

Typestate, as a type system extension for statically
verifying dynamically changing properties, was proposed in
\cite{StromYemini86Typestate, StromYellin93ExtendingTypestate}.
Aliasing causes problems for typestate-based systems because
the declared typestates of all aliases must
change whenever the state of the referred object changes.
Faced with the complexity of aliasing,
\cite{StromYemini86Typestate} resorted to a more controlled
language model which avoids aliasing. More recently proposed
typestate approaches use linear types for heap references to support 
state changes of dynamic allocated objects without 
addressing aliasing issues~\cite{DeLineFahndrich01EnforcingHighLevelProtocols}.

Motivated by the need to enforce safety properties in
low-level software systems, \cite{SmithETAL00AliasTypes,
WalkerMorrisett00AliasTypesRecursive,
CraryETAL99CalculusCapabilities} use extensions of linear
types to describe aliasing of objects and rely on language
design to avoid non-local type inference.  These systems take
a {\em construction based approach} that specifies data
structures as unfoldings of basic elaboration
steps~\cite{WalkerMorrisett00AliasTypesRecursive}.
Similarly to shape types
\cite{FradetMetayer97ShapeTypes,FradetMetayer96StructuredGamma}
and graph types \cite{KlarlundSchwartzbach93GraphTypes,
Moeller01PALE}, this allows tree-like data structures to be
expressed more precisely than using our roles, but cannot
approximate data structures such as sparse matrices.  More
importantly, this approach makes it difficult to express
nodes that are members of multiple data structures.
Handling multiple data structures is the essential
ingredient of our approach because the role of an object
depends on data structures in which it participates.

Like
shape analysis
techniques~\cite{ChaseETAL90AnalysisPointersStructures,
GhiyaHendren96TreeOrDag, SagivETAL96Destructive,
SagivETAL99Parametric} we have therefore adopted the {\em
constraint based approach} which characterizes data
structures in terms of the constraints that they
satisfy. 
The constraint based approach allows us to handle a wider
range of data structure while giving up some precision.
Like
\cite{XuETAL00SafetyCheckingPLDI,
XuETAL01TypestateCheckingMachineCode} we perform non-local
inference of program properties, but while
\cite{XuETAL00SafetyCheckingPLDI,
XuETAL01TypestateCheckingMachineCode} focus on linear
integer constraints and handle recursive data structures
conservatively, we do not handle integer arithmetic but have
a more precise representation of the heap.  At a higher
level, these approaches all focus on detailed properties of
individual data structures. We view our research as focusing
more on global aspects such as the participation of objects
in multiple data structures.

The path matrix approaches
\cite{GhiyaHendren95ConnectionAnalysis,
GhiyaHendren96TreeOrDag} have been used to implement
efficient interprocedural analyses that infer one level of
referencing relationships, but are not sufficiently precise
to track must aliases of heap objects for programs with
destructive updates of more complex data structures.

The use of the instantiation relation in role analysis
is analogous to the materialization operation of 
\cite{SagivETAL96Destructive, SagivETAL99Parametric}.
Role analysis can also track reachability properties, but we
use an abstraction relation based on graph homomorphism
rather than 3-valued logic.  Our split operation achieves a
similar goal to the focus operation of
\cite{SagivETAL99Parametric}.  However, the generic focus
algorithm of
\cite{LevAmi00TVLA} cannot handle the reachability predicate
which is needed for our split operation.  This is because it
conservatively refuses to focus on edges between two summary
nodes to avoid generating an infinite number of structures.
Rather than requiring definite values for reachability
predicate, our role analysis splits by reachability
properties in the abstract role graph, which illustrates the
flexibility of the homomorphism-based abstraction relation.
Another difference with \cite{SagivETAL99Parametric} is that
our role analysis does not require the developer to supply
the predicate update formulae for instrumentation
predicates.

A precise interprocedural
analysis~\cite{RinetzkySagiv01InterprocedualShapeAnalysis}
extends shape analysis techniques to treat activation
records as dynamically allocated structures.  The approach
also effectively synthesizes an application-specific set of
contexts. Our approach differs in that it uses a less
precise but more scalable treatment of procedures. It also
uses a compositional approach that analyzes each procedure
once to verify that it conforms to its specification. Like
\cite{XuETAL01TypestateCheckingMachineCode} our
interprocedural analysis can apply both may and must
effects, but our contexts are general graphs with summary
nodes and not trees.

Roles are similar to the ADDS and ASAP data structure
description languages
\cite{HummelETAL93ADDS, HummelETAL94ASAP,
HendrenETAL94GeneralDataDependence}.  These systems use
sound techniques to apply the data structure invariants for
parallelization and general dependence testing but do not
verify that the data structure invariants are preserved by
destructive updates of data structures \cite{Hummel98PhD}.

The object-oriented community has long been aware of
benefits that dynamically changing classes give in large
systems~\cite{Reenskaug96WorkingWithObjects}.  Recognizing
these benefits, researchers have proposed dynamic techniques
that change the class of an object
to reflect its state
changes~\cite{GammaETAL94DesignPatterns, GottlobETAL94Roles,
Chambers93PredicateClasses,
DrossopoulouETAL01Reclassification}.  These systems
illustrate the need for a static system that can verify the
correct use of objects with changing roles.

\nocite{JouvelotGifford91ReconstructionEffects, 
GhiyaHendren98PuttingPointerAnalysisWork,
IshtiaqOHearn01BIAssertionLanguage,
Deutsch94InterproceduralBeyond,
DiwanETAL98TypeBasedAliasAnalysis,
GuyerLin99AnnotationLanguage,
PlevyakETAL93DynamicStructures,
ChatterjeeETAL99RelevantContextInference,
ChengHwu00ModularInterproceduralPointerAnalysis,
CartwrightFagan91SoftTyping}

\section{Conclusion}

This paper proposes two key ideas: aliasing relationships
should determine, in large part, the state of each object,
and the type system should use the resulting object states
as its fundamental abstraction for describing procedure
interfaces and object referencing relationships. We present
a role system that realizes these two key ideas in a
concrete system, and present an analysis algorithm that can
verify that the program correctly respects the constraints
of this role system.  The result is that programmers can use
roles for a variety of purposes: to ensure the correctness
of extended procedure interfaces that take the roles of
parameters into account, to verify important data structure
consistency properties, to express how procedures move
objects between data structures, and to check that the
program correctly implements correlated relationships
between the states of multiple objects. We therefore expect
roles to improve the reliability of the program and its
transparency to developers and maintainers.

{\em Note. This November 2001 version of the technical
report corrects some errors in the formal description of the
analysis algorithm of the original technical report from
July 2001.}


\begin{figure*}
%
\begin{squeeze}
\[\begin{array}{rl}
\tu{\tu{H,\rho,K,\tau,E},\mu} \doeffects \tu{\Gb,\mu}
\mbox{ where }
& \tau[\mu^{-1}[\procread(\abstproc')]] \not\subseteq \procread(\abstproc) \\
\end{array}\]
\[\begin{array}{rl}
\tu{\tu{H,\rho,K,\tau,E},\mu} \doeffects G_t
\mbox{ where }
& \tau[\mu^{-1}[\procread(\abstproc')]] \subseteq \procread(\abstproc) \\
& \tu{H,\rho,K,\tau,E} \appeff{n_1,f_1} G_1 \appeff{} \cdots
                        \appeff{n_t,f_t} G_t \\
\end{array}\]
\[
   S = \{ \tu{n,f,n'} \in H \mid \tu{\mu(n),f,\mu(n')} \in \procmaywrite(\abstproc')
                      \cup \cup_i \procmustwrite_i(\abstproc') \}
\]
\[ 
   \{ \tu{n_1,f_1}, \ldots, \tu{n_t,f_t} \} = \{ \tu{n,f} \mid \tu{n,f,n'} \in S \}
\]
\centerline{}
\centerline{Single Write Effect Instantiation:}
\[
    \tu{H_1,\rho_1,K_1,\tau_1,E_1} \appeff{n,f} G'
\]
iff
\[\begin{array}{c|c|c}
\mbox{case}
&
\mbox{condition}
&
\mbox{result}
\\ \hline
\mbox{deterministic effect}
&
\vstack{
    \{ n_1 \mid \tu{n,f,n_1} \in S \} = \{ n_0 \} \mbox{ and }\\
    \exists i : \tu{\mu(n),f,\mu(n_0)} \in \procmustwrite_i(\abstproc') \\
}
&
\vstack{
          G' = \tu{H_2,\rho_1,K_1,\tau_1,E_2} \\
          H_2 = H_1 \setminus \{ \tu{n,f,n_1} \mid \tu{n,f,n_1} \in H_1 \} \\
           \qquad \qquad         \cup \{ \tu{n,f,n_0} \} \\
          E_2 = \updateMustWrite(E_1, \tu{\tau(n),f,\tau(n_0)}) \\
       }
\\ \hline
\vstack{\mbox{nondeterministic effect} \\
        \mbox{for non-parameters}}
&
\vstack{
    |\{ n_1 \mid \tu{n,f,n_1} \in S \}| > 1 \mbox{ or }\\
    \exists n_1 : \tu{\mu(n),f,\mu(n_1)} \in \procmaywrite(\abstproc') \\
  n \in \offstage(H) \\
  \{ \tu{\tau(n),f,\tau(n_1)} \mid \tu{n,f,n_1} \in S \} 
        \subseteq \procmaywrite(\abstproc) \\
}
&
\vstack{
         G' = \tu{H_2,\rho_1,K_1,\tau_1,E_2} \\
         H_2 = \removeIfMustWrite(H_1) \cup \\
         \qquad \qquad \{ \tu{n,f,n_1} \mid \tu{n,f,n_1} \in S \}
       }
\\ \hline
&
\vstack{
    |\{ n_1 \mid \tu{n,f,n_1} \in S \}| > 1 \mbox{ or }\\
    \exists n_1 : \tu{\mu(n),f,\mu(n_1)} \in \procmaywrite(\abstproc') \\
  n \in \offstage(H) \\
  \{ \tu{\tau(n),f,\tau(n_1)} \mid \tu{n,f,n_1} \in S \} 
        \not\subseteq \procmaywrite(\abstproc) \\
}
&
\vstack{
         G' = \Gb \\
       }
\\ \hline
\vstack{\mbox{nondeterministic effect} \\
        \mbox{for parameters}}
&
\vstack{
    |\{ n_1 \mid \tu{n,f,n_1} \in S \}| > 1 \mbox{ or }\\
    \exists n_1 : \tu{\mu(n),f,\mu(n_1)} \in \procmaywrite(\abstproc') \\
  n \notin \offstage(H) \\
  \{ \tu{\tau(n),f,\tau(n_1)} \mid \tu{n,f,n_1} \in S \} 
        \subseteq \procmaywrite(\abstproc) \\
}
&
\vstack{
         G' = \tu{H_2,\rho_1,K_1,\tau_1,E_2} \\
         H_0 = H_1 \setminus \{ \tu{n,f,n_1} \mid \tu{n,f,n_1} \in H_1 \} \\
         H_2 = H_1 \mbox{ or } H_2 = H_0 \cup
           \{ \tu{n,f,n_1} \} \\
         \tu{n,f,n_1} \in S
       }
\\ \hline
&
\vstack{
  \lnot(
    \{ n_1 \mid \tu{n,f,n_1} \in S \} = \{ n_1 \} \mbox{ and }\\
    \exists i : \tu{\mu(n),f,\mu(n_0)} \in \procmustwrite_i(\abstproc')) \\
  n \notin \offstage(H) \\
  \{ \tu{\tau(n),f,\tau(n_1)} \mid \tu{n,f,n_1} \in S \} 
        \not\subseteq \procmaywrite(\abstproc) \\
}
&
\vstack{
         G' = \Gb \\
       }
\\ \hline
\end{array}\]
\[
   \removeIfMustWrite(H_1) = \left\{\begin{array}{r@{,\>}l}
      H_1 \setminus \{ \tu{n,f,n'} \mid \tu{n,f,n'} \in H_1 \}
   & \mbox{ if } \exists i \> \exists n' : \tu{\mu(n),f,\mu(n')} \in \procmustwrite_i(\abstproc') \\
      H_1
   & \mbox{ otherwise} \\           \end{array}\right.
\]
\end{squeeze}
\caption{Effect Instantiation}
\label{fig:effectInst}
\end{figure*}



\bibliographystyle{plain}
\bibliography{pnew}

\end{document}